%% file: TRK-17-001_temp.tex
\pdfoutput=1

\documentclass[11pt,twoside,a4paper,cmspaper,final,collab]{cms-tdr}
\def\svnVersion{479855P}\def\cmsCernNoTag{CERN-EP-2018-144}\def\cmsCernDate{\today}\def\cmsMessage{Published in the Journal of Instrumentation as \href{http://dx.doi.org/10.1088/1748-0221/13/10/P10034}{\doi{10.1088/1748-0221/13/10/P10034.}}}

\begin{document}\cmsNoteHeader{TRK-17-001}

\hyphenation{had-ron-i-za-tion}
\hyphenation{cal-or-i-me-ter}
\hyphenation{de-vices}
\RCS$HeadURL: svn+ssh://svn.cern.ch/reps/tdr2/papers/TRK-17-001/trunk/TRK-17-001.tex $
\RCS$Id: TRK-17-001.tex 474956 2018-09-14 06:59:51Z kropiv $
\newlength\cmsFigWidth
\ifthenelse{\boolean{cms@external}}{\setlength\cmsFigWidth{0.85\columnwidth}}{\setlength\cmsFigWidth{0.4\textwidth}}
\ifthenelse{\boolean{cms@external}}{\providecommand{\cmsLeft}{top\xspace}}{\providecommand{\cmsLeft}{left\xspace}}
\ifthenelse{\boolean{cms@external}}{\providecommand{\cmsRight}{bottom\xspace}}{\providecommand{\cmsRight}{right\xspace}}
\ifthenelse{\boolean{cms@external}}{\providecommand{\CL}{C.L.\xspace}}{\providecommand{\CL}{CL\xspace}}
\providecommand{\NA}{\ensuremath{\text{---}}}
\newcommand{\rc}{\ensuremath{R}\xspace}
\newcommand{\Rnear}{\ensuremath{R^{\text{near}}}\xspace}
\newcommand{\Rfar}{\ensuremath{R^{\text{far}}}\xspace}
\newcommand{\Rx}{\ensuremath{R_{\mathrm{x}}}\xspace}
\newcommand{\Ry}{\ensuremath{R_{\mathrm{y}}}\xspace}
\newcommand{\xz}{\ensuremath{x_{0}}\xspace}
\newcommand{\yz}{\ensuremath{y_{0}}\xspace}
\newcommand{\xI}{\ensuremath{x_{\mathrm{i}}}\xspace}
\newcommand{\yI}{\ensuremath{y_{\mathrm{i}}}\xspace}
\newcommand{\xbs}{\ensuremath{x_{\text{bs}}}\xspace}
\newcommand{\ybs}{\ensuremath{y_{\text{bs}}}\xspace}
\newcommand{\yr}{\ensuremath{y}\xspace}
\newcommand{\rhomin}{\ensuremath{\rho_{\text{min}}}\xspace}
\newcommand{\rhomax}{\ensuremath{\rho_{\text{max}}}\xspace}
\newcommand{\rhoI}{\ensuremath{\rho_{\mathrm{i}}(\xz, \yz)}\xspace}
\newcommand{\rhoZero}{\ensuremath{\rho(\xz, \yz)}\xspace}
\newcommand{\chitwo}{\ensuremath{\chi^2}\xspace}
\newcommand{\Pc}{\ensuremath{P_{\mathrm{C}}}\xspace}
\newcommand{\PcOne}{\ensuremath{P_{\mathrm{C1}}}\xspace}
\newcommand{\PcTwo}{\ensuremath{P_{\mathrm{C2}}}\xspace}
\newcommand{\PcThree}{\ensuremath{P_{\mathrm{C3}}}\xspace}
\newcommand{\PV}{\ensuremath{P_{\mathrm{V}}}\xspace}
\newcommand{\PG}{\ensuremath{P_{\mathrm{G}}}\xspace}
\newcommand{\PGpr}{\ensuremath{P_{\mathrm{G}}'}\xspace}
\newcommand{\xZfar}{\ensuremath{\xz^{\text{far}}}\xspace}
\newcommand{\xZnear}{\ensuremath{\xz^{\text{near}}}\xspace}
\newcommand{\yZfar}{\ensuremath{\yz^{\text{far}}}\xspace}
\newcommand{\yZnear}{\ensuremath{\yz^{\text{near}}}\xspace}
\newcommand{\lambdaI}{\ensuremath{\lambda_{\mathrm{I}}}\xspace}
\newcommand{\nLost}{\ensuremath{n_{\text{lost}}}\xspace}
\newcommand{\dZero}{\ensuremath{d_0}\xspace}
\newcommand{\dm}{\ensuremath{d_{\mathrm{m}}}\xspace}
\newcommand{\AbsZbarrel}{\ensuremath{\abs{z} < 25\unit{cm}}\xspace}
\newcommand{\Fi}{\ensuremath{F_{\mathrm{i}}}\xspace}
\newcommand{\Fref}{\ensuremath{F_{\text{ref}}}\xspace}
\newcommand{\Ni}{\ensuremath{N_{\mathrm{i}}}\xspace}
\newcommand{\Bi}{\ensuremath{B_{\mathrm{i}}}\xspace}
\newcommand{\nSigma}{\ensuremath{n_{\sigma}}\xspace}
\newcommand{\sigmaNI}{\ensuremath{\sigma_{\text{NI}}}\xspace}

\cmsNoteHeader{TRK-17-001}
\title{Precision measurement of the structure of the CMS inner tracking system using nuclear interactions}

\date{\today}

\abstract{
The structure of the CMS inner tracking system has been studied using nuclear interactions of hadrons striking its material.
Data from proton-proton collisions at a center-of-mass energy of 13\TeV recorded in 2015 at the LHC are used to reconstruct millions of secondary vertices from these nuclear interactions. Precise positions of the beam pipe and the inner tracking system elements, such as the pixel detector support tube, and barrel pixel detector inner shield and support rails, are determined using these vertices. These measurements are important for detector simulations, detector upgrades, and to identify any changes in the positions of inactive elements.
}
\hypersetup{
pdfauthor={CMS Collaboration},
pdftitle={Precision measurement of the structure of the CMS inner tracking system using nuclear interactions},
pdfsubject={CMS},
pdfkeywords={CMS, physics, tracker}}

\maketitle

\section{Introduction}

Precision mapping of the material within the tracking volume of the CMS detector~\cite{Chatrchyan:2008zzk}
is important for the experiment's measurement goals.
The material affects the
reconstruction of events through multiple
scattering, energy loss, electron bremsstrahlung, photon conversions,
and nuclear interactions (NIs), of the particles produced in proton-proton collisions.
The analysis presented here uses reconstructed NIs to
precisely measure the positions of inactive elements surrounding
the proton-proton collision point, such as the beam pipe and the inner mechanical structures
of the pixel detector. This information is needed to validate simulations of the CMS detector and
to identify any shifts in the positions of inactive elements.
It can also be used in searches for long-lived particles with displaced vertices~\cite{Khachatryan:2016sfv,Sirunyan:2017jdo}.

An accurate simulation of the effects of inactive material is necessary for a
proper reconstruction of all particles produced in a proton-proton collision event.
In particular, the material
closest to the interaction region affects the track position
resolution, which, in turn, strongly affects
the b tagging performance~\cite{Sirunyan:2017ezt}.
Substantial effort has been invested into the implementation of the detailed
\GEANTfour~\cite{Agostinelli:2002hh, GEANT}
geometry used to simulate the detector response.  Previous studies have
addressed the systematic uncertainties related to the tracker material
simulation~\cite{Migliore:2010cva}, validation of the
simulation with early data~\cite{CMS-PAS-TRK-10-003}, and more
accurate calibrations based on the data to improve the resolution of
calorimeter-based observables~\cite{Khachatryan:2015iwa}.

Identifying shifts in the positions of inactive elements is important not only for accurate detector response simulation,
but also for CMS tracker upgrade designs.
Since the pixel detectors are installed
with the beam pipe
already in place, an accurate measurement of the beam pipe position
is of paramount importance. As a
consequence of the beam pipe mechanical characteristics and support
structure design,
its final position can be different
from the nominal one at the millimeter level~\cite{Dominguez:1481838}.
For the design of the Phase-1 upgrade of the pixel detector, which was installed in Spring 2017,
NI imaging was used to conclude that the sagging of the beam pipe between the supports
was small enough to be of no concern.
Both the original and new versions of the CMS  pixel detector are split into two half-cylinders
and inserted by sliding these two halves into place
by means of appropriate rails.
This installation method does not
provide accuracy and reproducibility of the pixel detector positioning
below the level of a few millimeters. The evaluation and understanding of
these position uncertainties, which are comparable with
mechanical clearances between the pixel
detector and the beam pipe, were important inputs for the design of
the new support system and helped to establish reliable installation procedures
for the Phase-1 upgrade of the pixel detector.
Clearances may change during detector operation because of deformation due to gravity, and variations in
vacuum pressure, temperature, and magnetic field~\cite{Dominguez:1481838}.
The innermost layer of the Phase-1 upgrade of the pixel detector is even
closer to the beam pipe than the previous innermost pixel layer~\cite{Dominguez:1481838},
but the clearances were well understood from the NI imaging of the pixel detector support tube.

Nuclear interaction reconstruction has
been developed in the past~\cite{CMS-PAS-TRK-10-003, Aad:2011cxa, Aaboud:2016poq, Aaboud:2017pjd} as a powerful
tool for investigating inactive material in a tracking detector.
Reconstructed NIs profit from higher multiplicity and larger scattering angles of secondary tracks 
emerging from the NI vertex, resulting in better vertex position resolution along the direction 
of the impinging particle compared to photon conversion vertices~\cite{CMS-PAS-TRK-10-003}. 
Thus, the vertex resolution of reconstructed NIs is typically sub-millimeter, and the large number of NIs leads to a precision
of the order of 100\unit{\mum} in determining the positions of inactive elements of the detector.
The position resolution has negligible statistical uncertainties and is limited by systematic uncertainties.

In the alignment procedure, reconstructed tracks are used to measure the positions of the pixel detector layers
relative to the outer tracking system~\cite{Chatrchyan:2014wfa}. However, the only way to accurately
measure the positions of the inactive elements (such as the beam pipe) with respect
to the tracking detectors is to use NIs. This paper
describes their effective use for a post-installation survey of the critical detector
region surrounding the interaction point.

The paper is organized as follows.
In Section~\ref{sec:CMSdetector} a brief description of the CMS detector and its coordinate
system is given. Section~\ref{sec:NI_rec} summarizes the data sets used in the analysis
and the reconstruction method for the NIs is
presented. Section~\ref{sec:Methodology} describes the position measurement method, while
in Section~\ref{sec:InnerTracker} the actual measurements for the original pixel detector are presented.
In Section~\ref{sec:Systematics} the systematic uncertainties are addressed.
In Section~\ref{sec:Survey} the comparison of these measurements with technical surveys is discussed,
followed by a summary in Section~\ref{sec:Conclusions}.

\section {CMS detector}
\label{sec:CMSdetector}

The CMS detector is one of
two general-purpose detectors operating at the
LHC facility at CERN. One of the central features of the CMS
detector is
a superconducting solenoid of 6\unit{m} internal diameter, providing a magnetic field of 3.8\unit{T},
which enables the measurement of charged particle momenta
by reconstructing their trajectories as they traverse the CMS
tracking system. The CMS experiment uses a right-handed coordinate system, with the
origin at the nominal interaction point, the $x$ axis pointing to the
center of the LHC ring, the $y$ axis pointing up (perpendicular to the
LHC plane), and the $z$ axis along the counterclockwise beam direction. The
azimuthal angle $\phi$ is measured in the $x$-$y$ plane, with $\phi=0$ along the
positive $x$ axis, and $\phi=\pi/2$ along the positive $y$ axis. The radial
coordinate in this plane is denoted by $r$.

\begin{figure}[t]
\centering
\includegraphics[width=0.9\textwidth]{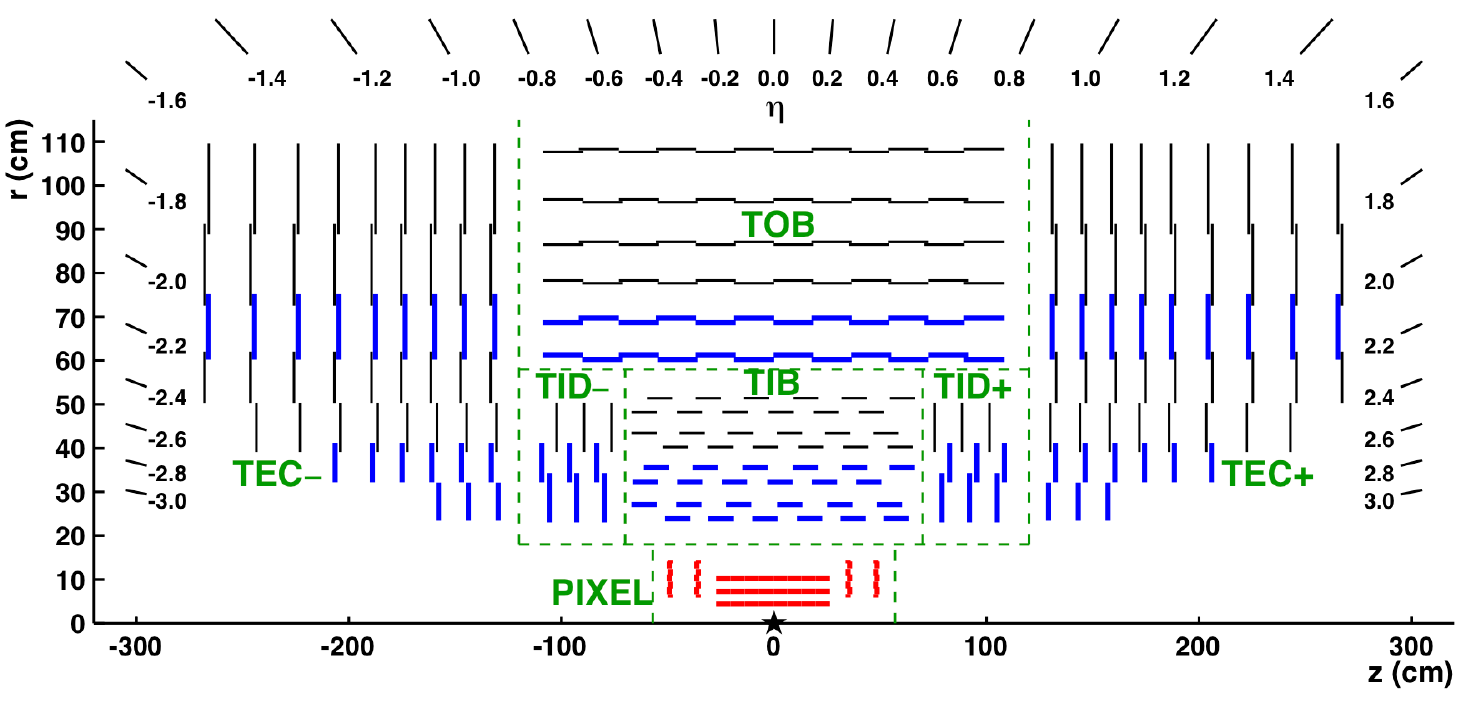}\\
\includegraphics[width=0.7\textwidth]{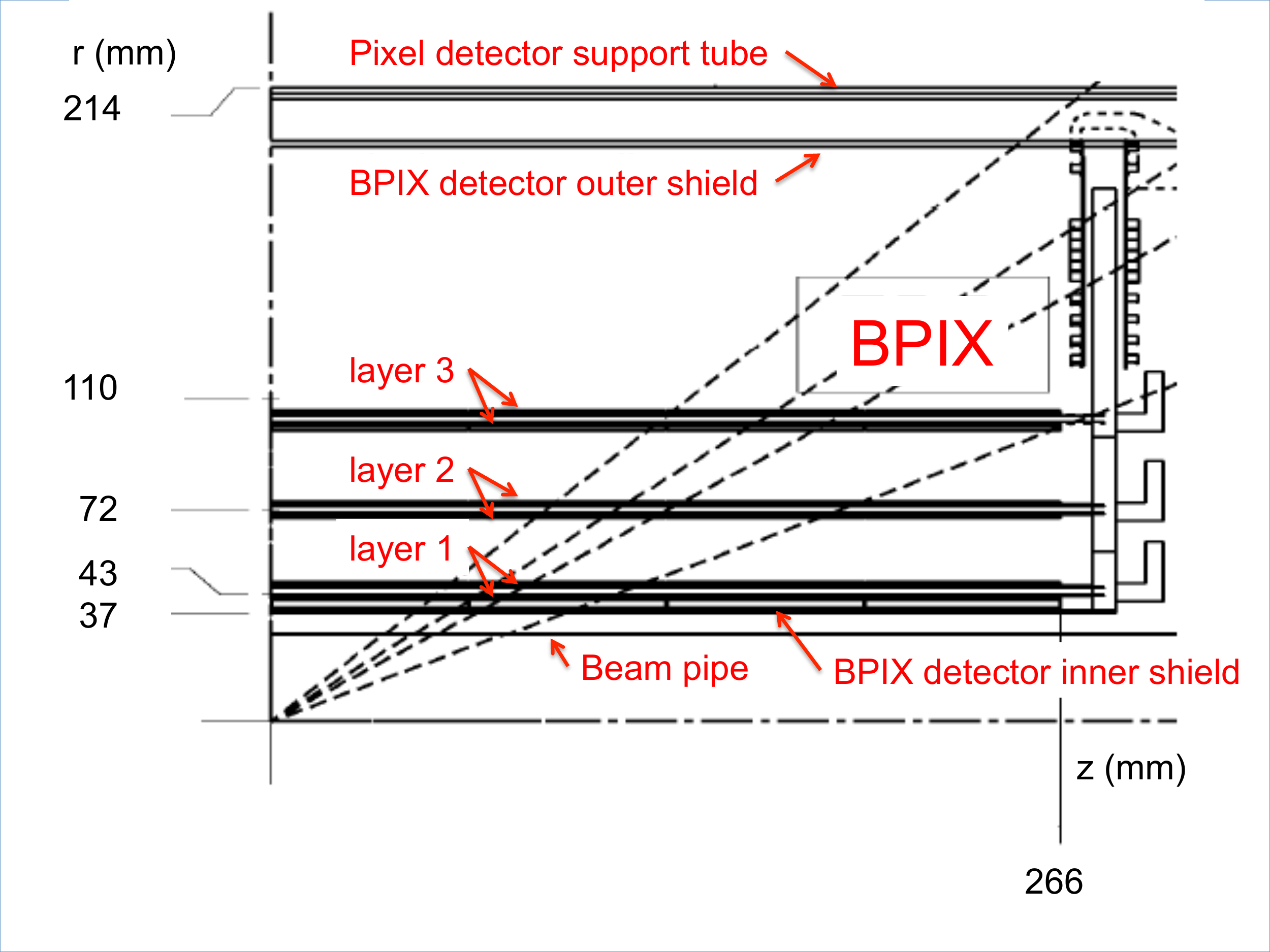}\\
\caption{(upper) Schematic view of the CMS tracking detector~\cite{Chatrchyan:2008zzk},
and (lower) closeup view of the region around the original BPIX detector with labels
identifying
pixel detector support tube, BPIX detector outer and inner shields, three BPIX detector layers, and beam pipe.
}
\label{fig:tklayout}
\end{figure}

The CMS tracking system, shown in Fig.~\ref{fig:tklayout}\,(upper), consists of two
main detectors: the smaller inner pixel detector and the larger silicon strip detector.
The original pixel detector had three barrel pixel (BPIX) layers and two endcap disks per side, covering the region from 4 to 15\unit{cm} in
radius, and spanning 98\unit{cm}
along the LHC beam axis. The silicon strip tracking system has ten barrel layers and twelve endcap disks per side, covering the region from 25 to 110\unit{cm} in
radius, and spanning 560\unit{cm}
along the LHC beam axis. The tracking system acceptance extends up to a
pseudorapidity of $\abs{\eta}=2.5$. The silicon strip tracking system has four subsystems.
The innermost four barrel layers comprise the tracker inner barrel (TIB) detector,
and the outer six barrel layers form the tracker outer barrel (TOB) detector. The three endcap disks
to either side of the TIB detector form the tracker inner disks (TID$-$ and TID$+$),
and the nine endcap disks at each end constitute the tracker endcap (TEC$-$ and TEC$+$).

\begin{figure}[t]
\centering
\includegraphics[width=0.6\textwidth]{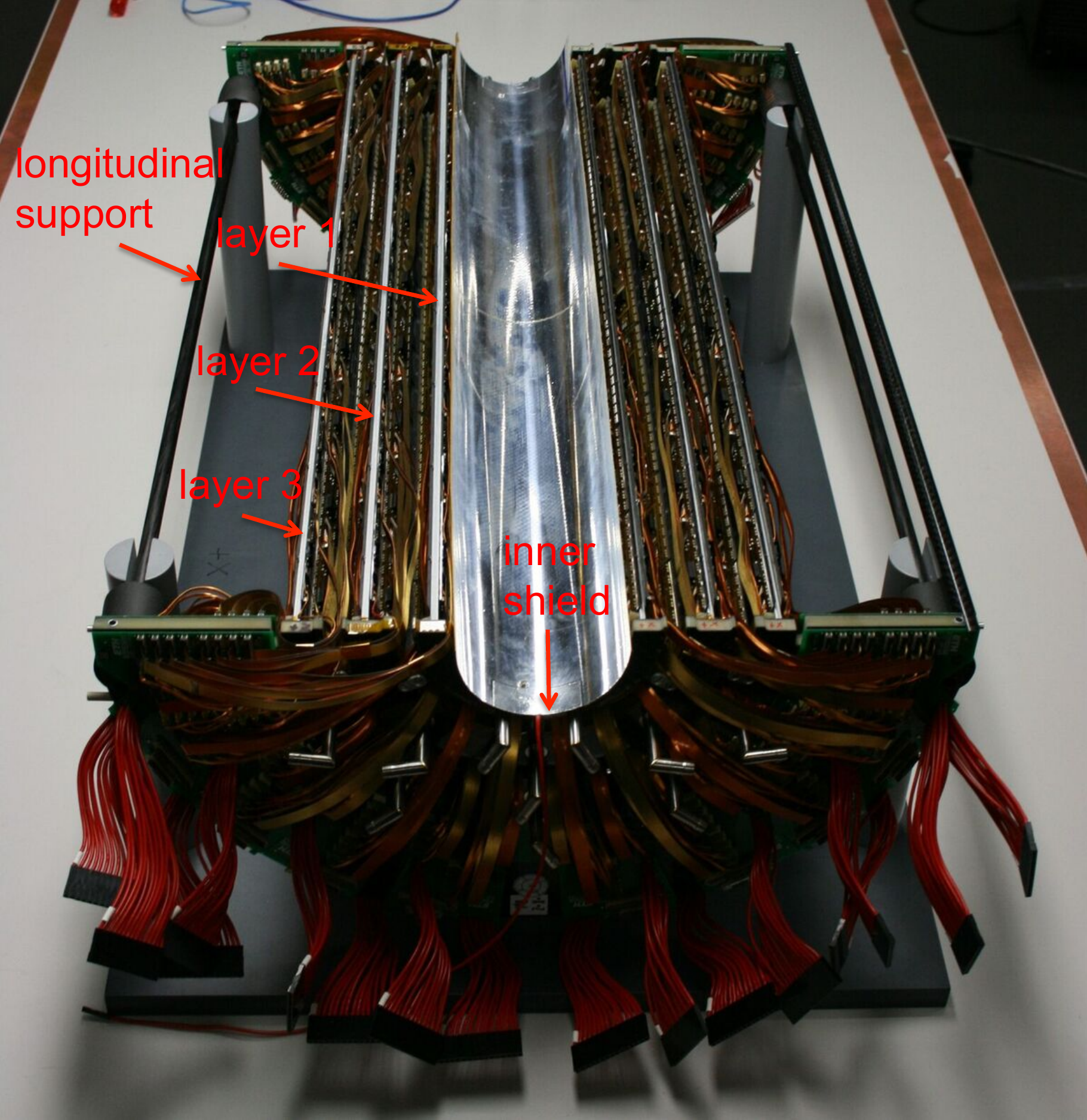}
\includegraphics[width=0.3275\textwidth]{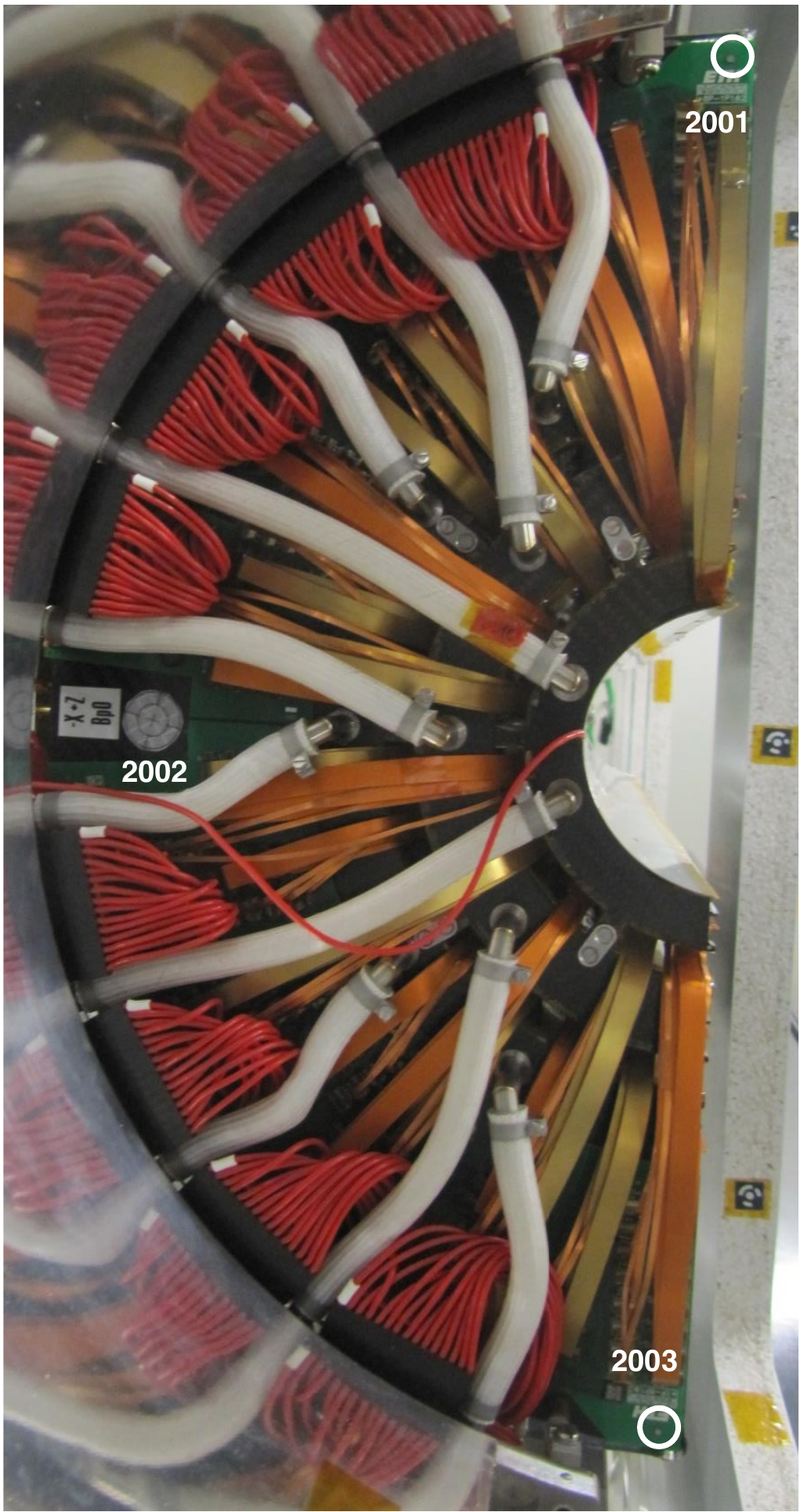}
\caption{
(left) Photograph of one half of the BPIX detector showing
longitudinal support, three layers, and inner shield.
(right) Photograph showing an end of the BPIX detector while standing on the installation cassette.
Optical targets, indicated by the numbers 2001, 2002, and 2003, are used to locate the BPIX detector within the CMS cavern.
Photographs by Antje Behrens, CERN.}
\label{fig:PhotoHalve}
\end{figure}

The particular structural elements studied in this paper are the inactive elements that surround the BPIX detector,
shown in Fig.~\ref{fig:tklayout}\,(lower):
the pixel detector support tube, the BPIX detector outer and inner shields, three BPIX detector layers, and the beam pipe.
The beam pipe is the innermost structure and, proceeding outward, the next structure
is the BPIX detector inner shield.
The  BPIX detector and its inner shield are composed of two semi-circular halves, which are called
`far' for the structure outside the LHC ring ($x < 0$)
and `near' for the structure inside of the LHC ring ($x > 0$).
Photographs of one of the halves of the BPIX detector are shown in
Fig.~\ref{fig:PhotoHalve}.
The BPIX detector support rails are located on the outer edge of the BPIX detector,
and  hold its layers in place. The pixel detector support tube encloses the BPIX detector and
the support rails.

\section{Data sample and nuclear interaction reconstruction}\label{sec:NI_rec}

The data set used in this analysis was recorded in 2015 from proton-proton collisions at a center-of-mass energy of 13\TeV at the LHC, and
corresponds to an integrated luminosity of 2.5\fbinv.
Studies were also performed using Monte Carlo (MC) simulations
with \PYTHIA~8~\cite{Pythia8,Sjostrand:2014zea,Khachatryan:2015pea}
based on single charged pions generated uniformly in $\eta$
and $\phi$ at different fixed momenta.
The resulting single-pion samples are processed through a \GEANTfour-based
detector simulation.

The data sample was selected using two main criteria.
First, the density of particles should not be too large
because tracks coming from the primary interaction may be
mismeasured and the NI reconstruction algorithm may assign them to an NI vertex.
These random combinations of primary tracks are the main source of background, and what we label ``misreconstructed'' NIs.
Second, the sample should have a sufficient number of events to compensate for the low efficiency of the NI reconstruction.
The set of events obtained using a collection of triggers~\cite{Khachatryan:2016bia}
that require at least one high \pt muon fulfills both of these criteria since these events tend to 
have fewer areas of high-density hadronic activity than events triggered only by jets.

Previous studies show that the overall material thickness of the silicon tracking system varies between 0.1--0.5~\lambdaI, where \lambdaI is the
characteristic nuclear interaction length~\cite{CMS_CR_2008_007}. Simulations show that approximately 5\%
of charged pions with transverse momentum $\pt\approx5\GeV$
interact in the tracking system~\cite{CMS-PAS-TRK-10-003}.
Each NI can create a displaced vertex within the tracking volume, with an incoming particle and a few outgoing particles.
In this analysis we look for NI vertices that have at least three associated charged particles, as discussed in more detail below.

In the methodology used to reconstruct NI vertices, the first step is to find the tracks using the CMS iterative
tracking algorithm~\cite{Chatrchyan:2014fea, Sirunyan:2017ulk}.
This algorithm proceeds with a sequence of ten iterations. For each iteration, a specific seeding pattern is identified requiring two or three hits from pixel detector layers or strip detector stereo layers~\cite{Chatrchyan:2014fea}. 
Those seeds are forward propagated within the tracking volume and the tracks are retained
if quantities such as the total number of hits, \pt, quality of the fit, the transverse impact parameter with respect to the
primary vertex, \dZero, and the number of missing hits, \nLost, fulfill certain quality criteria.
This last variable is obtained by extrapolating the track's trajectory outward toward the calorimeters and inward toward the beam axis. The value of \nLost is then the number of strip and pixel detector layers crossed by the trajectory that have no measured hits.

At the end of each iteration, the hits associated with the identified tracks are masked to reduce the combinatorics of the next step.
The highest-quality tracks
are identified in the earliest iterations, while subsequent iterations select tracks with lower quality and larger combinatorics.

Tracks considered for NI reconstruction benefit from all ten iterations and are required to have $\pt > 200\MeV$ to reduce the number of misreconstructed NIs. They are classified into three categories according to their position relative to the NI vertex:
\begin{itemize}
\item Incoming tracks: we require $\dZero < 0.2\unit{cm}$, and at least three hits, with at most one hit after the NI vertex.
\item Outgoing tracks: we require $\dZero > 0.2\unit{cm}$, at least six hits, with at most one hit before the NI vertex, and $\nLost < 10$.
\item Merged tracks: we require $\dZero < 0.2\unit{cm}$, at least four hits, with at least two hits before, and two after the NI vertex, and $\nLost < 10$.
\end{itemize}

For an NI in the strip detector, the incoming charged particles may be reconstructed as short tracks seeded from pixel detector triplets or pairs of hits.
For an NI in the pixel detector the incoming charged particle leaves too few hits to be reconstructed. It can happen, though, that
the hits from the incoming charged particle are assigned by the tracking algorithm to an outgoing track that is much longer.
In that case, a merged track is obtained. The tracking of the outgoing charged particles typically uses strip-only or mixed pixel-strip detector seeding.
Finally, in NI cases where most of the momentum
of the incoming charged hadron is transferred to an outgoing one, the trajectories of the incoming and outgoing tracks may be assigned by the tracking algorithm to a single merged track.

\begin{figure}[t]
\centering
\includegraphics[width=0.3\textwidth]{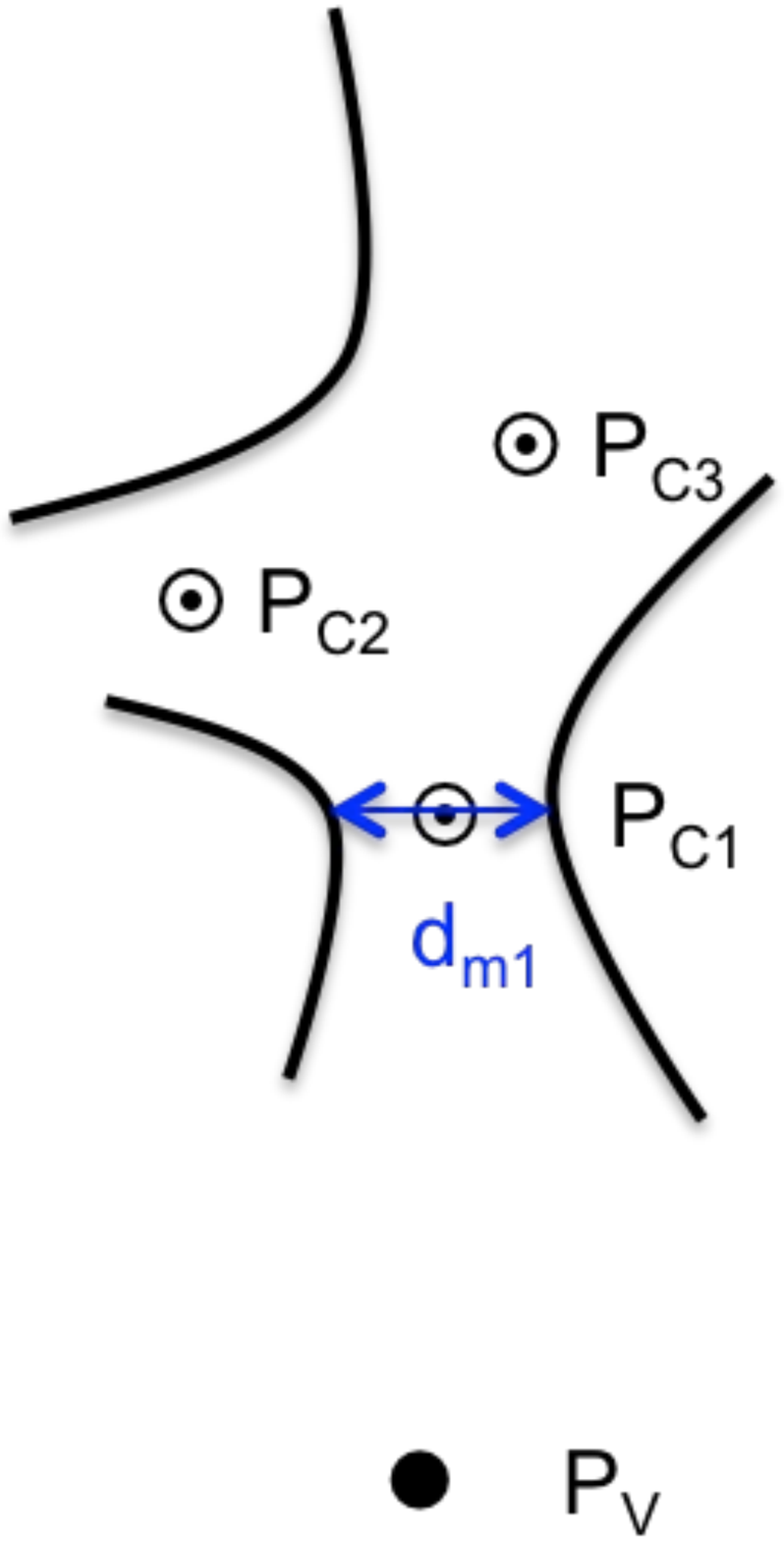}
\includegraphics[width=0.3\textwidth]{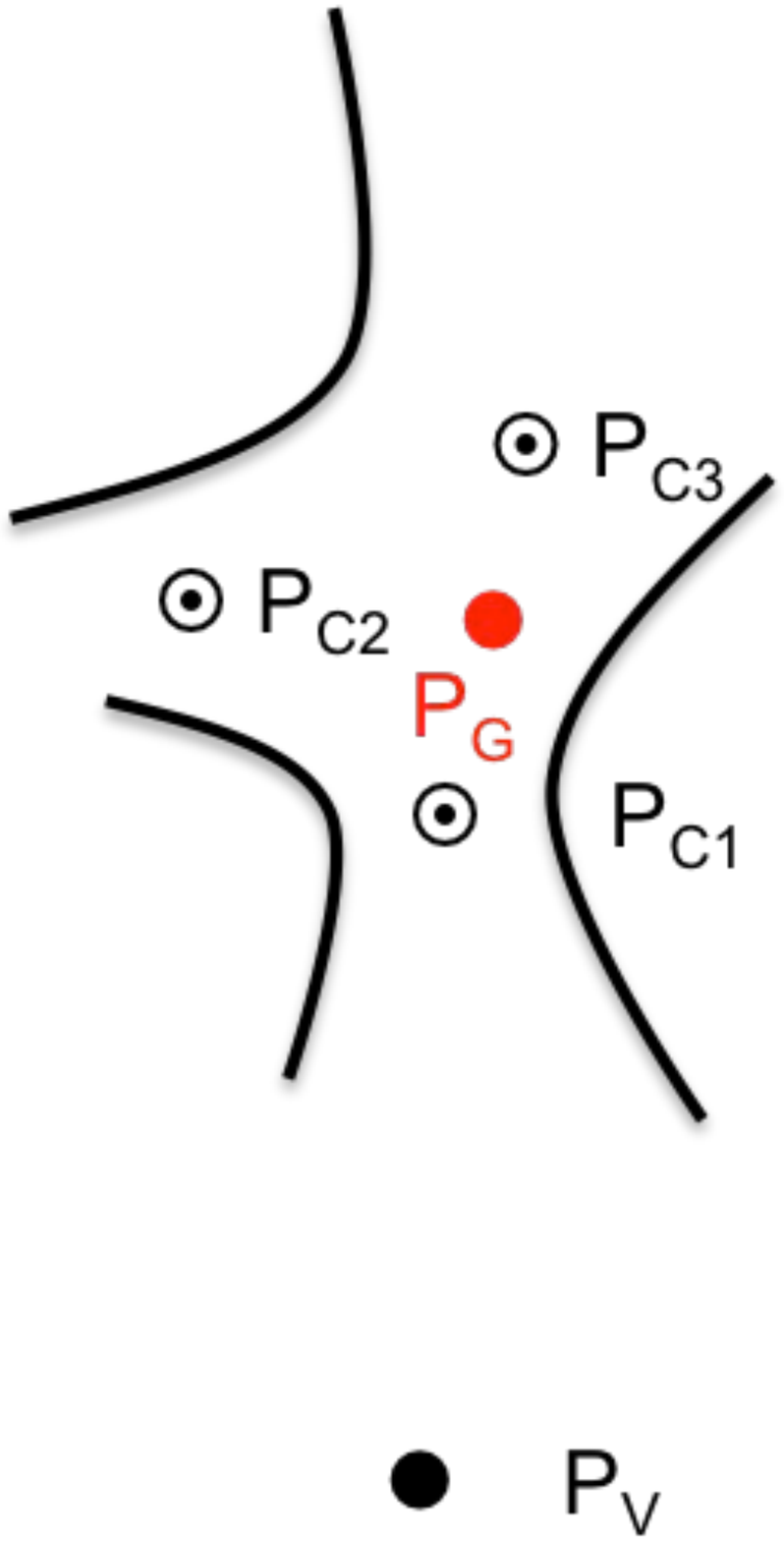}
\includegraphics[width=0.3\textwidth]{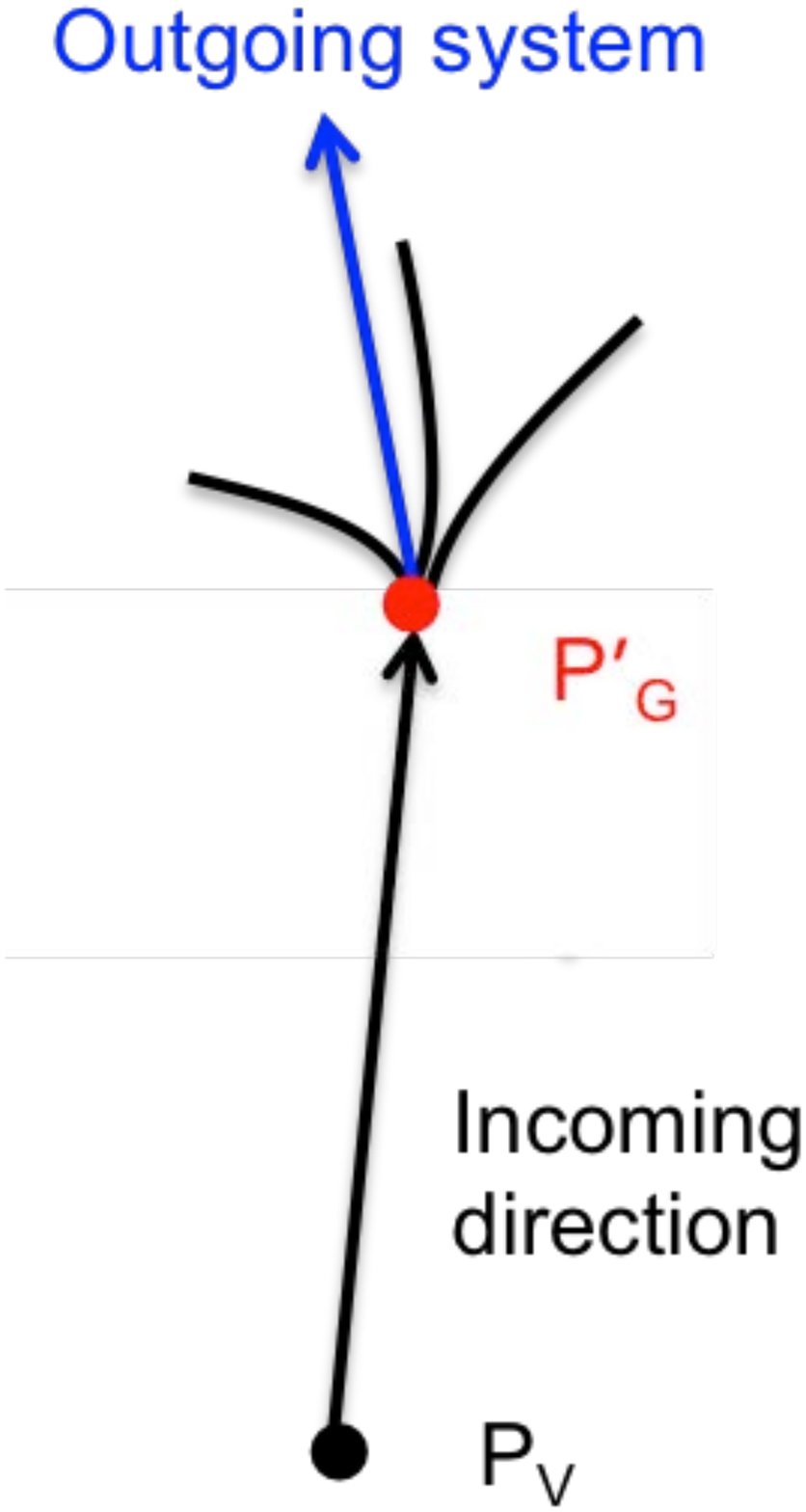}
\caption{
Schematic view of NI vertex reconstruction:
(left) a cluster of \Pc positions (\PcOne, \PcTwo, and \PcThree)
with the distance of closest approach \dm (labeled $d_{\mathrm{m1}}$), shown for \PcOne;
(center) the algorithm uses the three \Pc points to identify an aggregate position \PG;
(right) after refitting the track helices, the best vertex \PGpr is found
with indicated incoming direction from the primary vertex position, \PV, and outgoing system.
Black curves correspond to reconstructed charged particle tracks.
         }
\label{fig:SchematicView}
\end{figure}

A vertex reconstructed from the list of selected tracks identifies the position of the NI.
For each pair of tracks the points of closest approach are identified.
The length of the segment connecting these two points provides the distance of closest approach, \dm~\cite{Chatrchyan:2014fea}.
If $\dm < 0.5\unit{cm}$,
we consider the possibility that both tracks come from the same vertex.
The center, \Pc, of the segment is then considered as the best estimate of the position of this vertex.
This step is sketched in Fig.~\ref{fig:SchematicView} (left).

A three-dimensional (3D) clustering procedure is used to iteratively aggregate the \Pc positions of vertex candidates.
In practice, we start from the innermost \Pc (labeled \PcOne) with respect to the primary vertex position, \PV,  and look for the presence
of points within a cylinder of $\pm$5\unit{cm} along the direction of the vector $\overrightarrow{\PV \Pc\,}$ and 1\unit{cm} in radius.
The dimensions of this cylinder are defined to take into account the resolution of the track parameters.
If several points are found, the closest to \PcOne is selected, and
the barycenter position (\PG) between this point and \PcOne is calculated. The algorithm is iterated starting from \PG.
The search stops when no additional points are found within the cylinder.
This step is sketched in Fig.~\ref{fig:SchematicView} (center).
All points selected during the search are removed from the list of \Pc values and the algorithm is restarted.

Each \PG with its associated tracks is passed to the
adaptive vertex fitter (AVF)~\cite{Chatrchyan:2014fea}, which refits the helices of the tracks assuming a point close to \PG as the common vertex. An example is sketched in Fig.~\ref{fig:SchematicView} (right).
The AVF provides a list of vertex candidates with their best position estimates, \PGpr, as its output.

The set of outgoing tracks from a vertex candidate is referred to as the \textit{outgoing system}.
The Lorentz four-vectors of those tracks, assuming a pion mass for each track, define the kinematic properties of the outgoing system.
The incoming or merged track is referred to as the \textit{incoming system}. It provides the direction and kinematic properties of the impinging particle.
In case no incoming or merged track is present, the vector $\overrightarrow{\PV \PGpr\,}$ defines the incoming direction.

This list of vertex candidates is filtered with the following quality criteria designed to select NI vertices and reject conversions, decays in flight,
and misreconstructed NI vertices:

\begin{itemize}
\item At least three tracks are required, including incoming, merged, or outgoing tracks.
\item An NI candidate with more than one incoming and/or merged track is rejected.
\item The outgoing system must have an invariant mass of at least 1\GeV and $\pt > 500\MeV$.
\item The angle between the incoming and outgoing directions shall not exceed 15\de.
\item Vertex candidates located well inside the nominal beam pipe radius are removed since there is no material in that region.
\end{itemize}

With these criteria, 5.40 million events are found with at least one NI and these events yield a total of 5.71 million NIs.

The position resolution of the reconstructed NI vertices is estimated using the single-pion simulation.
The positions of the actual NI and its reconstruction are recorded and the differences are compared
in different regions of the detector. Within the beam pipe, the typical resolution perpendicular to the direction of the particle's
propagation is of the order of 50\unit{\mum}. The resolution degrades at larger radius due to the smaller number of pixel detector hits included
in the tracking. Within the pixel detector volume, the resolution is approximately 100\unit{\mum}, and it increases to 200\unit{\mum}
within the pixel detector support tube. The perpendicular vertex position resolution is the main factor in determining
how well the centers of the different structures can be located. The vertex position resolution along the direction of
propagation of the incoming track is worse than
in the perpendicular direction because the combining of the individual track locations is less precise in this direction.
The resolutions along the track direction are 300\unit{\mum} within the beam pipe, 500\unit{\mum} within the pixel detector, and
1000\unit{\mum} within the pixel detector support tube. These resolutions are smaller than the element thicknesses in the
different structures under consideration and the impact on the uncertainties associated with the measurement procedure remains limited.

The purity of the NI sample depends on the region under consideration and the signal-over-background ratio is about 1.5, 0.5, and 8 for the beam pipe,
BPIX detector inner shield, and pixel detector support tube, respectively.
The misreconstruction rate decreases as track density decreases and so it is smaller at higher radius.
The misreconstruction rate for each measurement is estimated from data by looking at a region to the side of the structure under
consideration, where no material is expected.

\begin{figure}[t]
\centering
\includegraphics[width=1.0\textwidth]{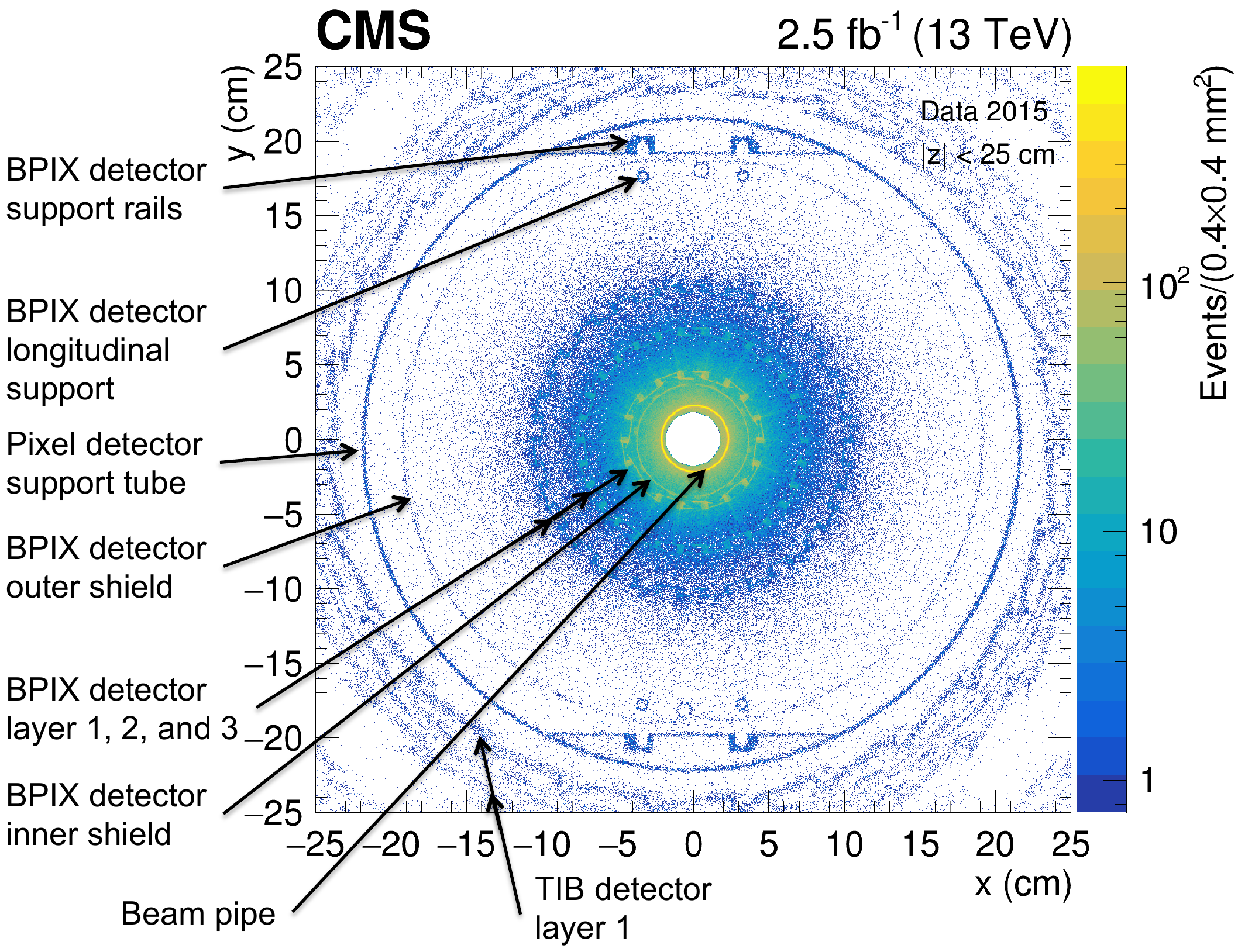}
\caption{
         Hadrography of the tracking system in the $x$-$y$ plane in the barrel region (\AbsZbarrel).
         The density of NI vertices is indicated by the color scale.
         The signatures of the beam pipe, the BPIX detector with its support, and the first layer of the TIB detector can be observed above the background of misreconstructed NIs.}
\label{fig:TrackerHadrography}
\end{figure}

The ``hadrography" in the $x$-$y$ plane of the tracking system in the barrel region (\AbsZbarrel) is provided in Fig.~\ref{fig:TrackerHadrography}.
The signatures of the beam pipe, the BPIX detector with its support, and the first layer of the TIB detector can be seen.

\section{Analysis method}\label{sec:Methodology}

In this analysis we focus on the measurement of the positions of the inactive elements within the inner tracking system.
All the inactive elements under consideration except for the support rails have a cylindrical geometry with their axes being
collinear to the beam axis.
For all the structures but the support rails, the axis position is within a few millimeters of $(0,0)$,
the origin of the CMS offline coordinate system, which is discussed in Section~\ref{CMScoordinate}.
By design, the thicknesses of the structures do not exceed a few millimeters to keep the amount of material within the inner tracking system to a minimum.
These properties of the components under consideration allow a significant simplification of the fitting technique.
Instead of a complex fit of a 3D structure, we fit the parameters of a function in the $x$-$y$ plane using shapes such as circles, half-circles, or ellipses.

The slight displacement of the structures' axes with respect to the beam axis induces differences in the azimuthal hadron fluxes seen
by different elements of the structure. We correct for that effect locally by reweighting events~\cite{CMS-PAS-TRK-10-003} using a
geometric factor for each bin i, \Fi, that accounts for the small flux effect. To a first approximation we can write
$\Fi = 1/r_{{\mathrm{i}}, \text{bs}}^2$, where
$r_{\text{bs}}$ is the radial distance calculated with respect to the average
over the data-taking period of the beam spot position that was computed
using the method from Ref.~\cite{Chatrchyan:2014fea}.
For the 2015 data-taking period, the average beam spot position in the transverse plane was
$\xbs =0.8\unit{mm}$ and $\ybs = 0.9\unit{mm}$~\cite{CMS-DP-2016-012}.
The beam spot position varied during the year by distances of less than 0.1\unit{mm}.

The measurement of each cylindrical structure follows the same steps:
\begin{enumerate}
\item The NI vertices are selected within a ring of a few centimeters around the structure and a binned position distribution in i
is made in the $x$-$y$ plane.
The chosen bin sizes are smaller
than the thicknesses of the objects being studied, but large enough to allow for stable fitting procedures.
The two-dimensional (2D) bin size in the $x$-$y$ plane used for the beam pipe and BPIX detector inner shield measurements is $500\times500\unit{\mum}^2$.
For the pixel detector support tube the bin size is $1700\times1700\unit{\mum}^2$, and for the BPIX detector support rails a bin size of $800\times800\unit{\mum}^2$ is used.
\item The resulting distribution is sliced into 40 regions in $\phi$. The slices where additional structures such as cooling pipes or collars, are visible near the main structure are rejected from the analysis.
\item In each slice, the 2D distribution is projected along the $\phi$ coordinate
and the distribution of \rhoI values is considered, where $\rhoI = \sqrt{\smash[b]{(\xI-\xz)^2+(\yI-\yz)^2}}$
is the radial position of the center of bin i in the relative coordinates of the structure, with origin $(\xz, \yz)$.
The signal region is defined using \rhomin and \rhomax values chosen around the expected position of the structure.
The combinatorial background is estimated using signal sidebands in \rhoI,
which are fit by an exponential function,
yielding a value \Bi of background events under the signal in the $x$-$y$ plane for each bin i.
\item A \chitwo is defined for a circular shape:
\begin{linenomath}
\begin{equation}
\chitwo = \sum_{{\mathrm{i}}:\rhomin < \rhoI < \rhomax}\frac{{\text{max}}[0,(\Ni - \Bi - \nSigma\sqrt{\Bi})]\, \Fi/\Fref \, (\rhoI - \rc)^2}{\sigmaNI^2},
\label{eq:chitwo}
\end{equation}
\end{linenomath}
where $\sigmaNI = 100\unit{\mum}$ is the typical NI vertex resolution, \Ni is the number of events in bin i,
\nSigma is the number of standard deviations above the nominal background,
and \rc is the radius of the structure.
 In the case of an ellipse, \rc becomes dependent on \xI and \yI through a relation parametrized
by the semi-minor axis (\Rx) and semi-major axis (\Ry).
The correction factor $\Fi/\Fref$ is defined to mitigate the small differences in hadron flux between bins.
To keep this factor as close as possible to 1, we take \Fref to be the value of the flux at the expected radius \rc
around the beam spot position. During the minimization procedure we do not recalculate \Fref. The overall impact of
the flux factor on the final result is small.
\item We subtract the mean background plus twice
the expected background uncertainty
(using $\nSigma=2$)
to maximize the signal purity and improve the visibility of the structures.
\item A minimization of the \chitwo is subsequently performed with \rc, \xz, and \yz, as free parameters.
\end{enumerate}

\section{Measurements of pixel detector positions}\label{sec:InnerTracker}

\subsection{Measurement of the beam pipe position}\label{sec:BeamPipe}

\begin{figure}
\centering
\includegraphics[width=0.6\textwidth]{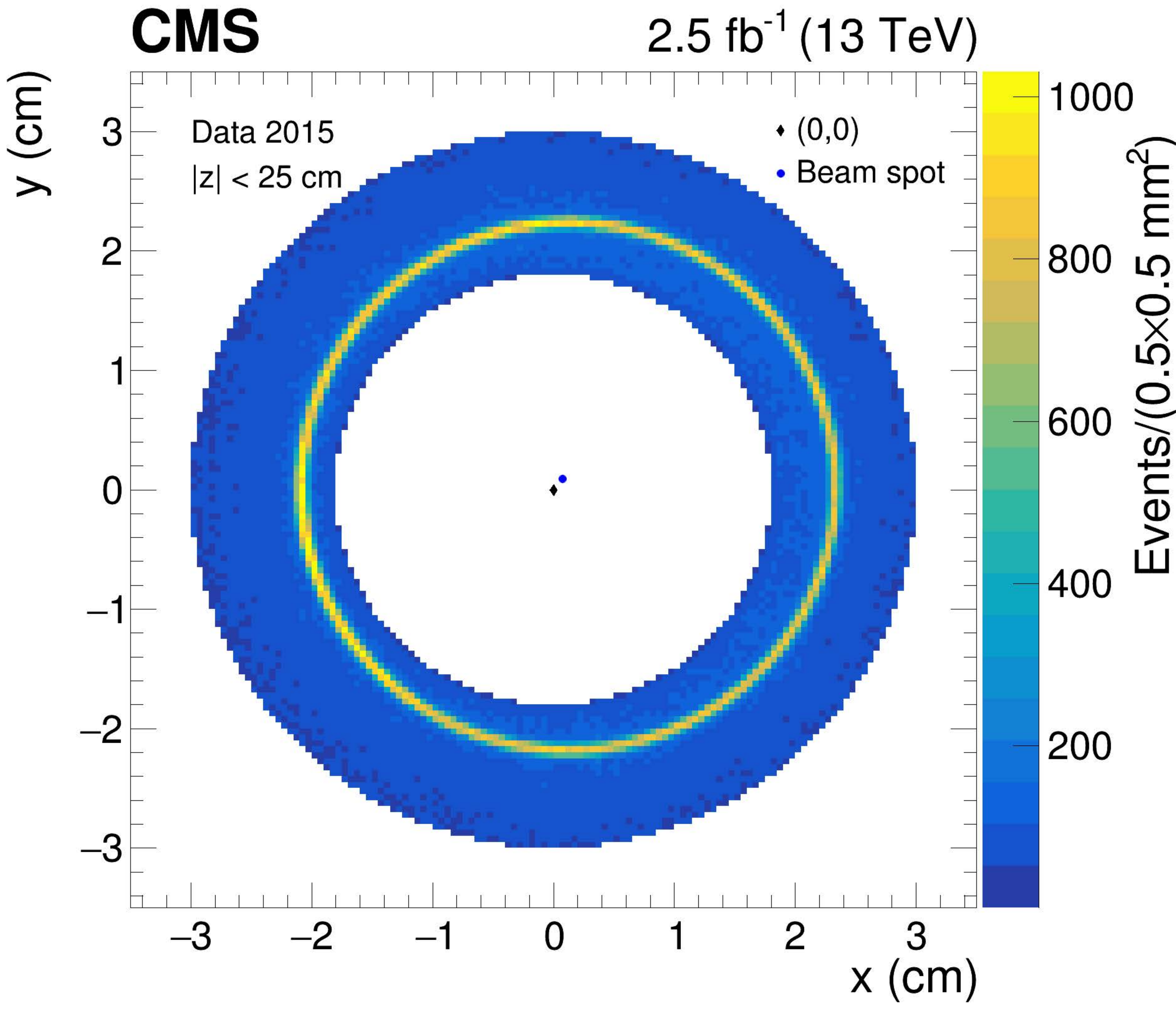}
\caption{The beam pipe region viewed in the $x$-$y$ plane for \AbsZbarrel before background subtraction.
The density of NI vertices is indicated by the color scale.
$(0,0)$ is the origin of the CMS offline coordinate system, which is discussed in Section~\ref{CMScoordinate}.
The blue point in the center of the distribution corresponds to the average beam spot position
of $\xbs=0.8\unit{mm}$ and $\ybs=0.9\unit{mm}$ in 2015.}
\label{fig:BeamPipe_Draw}
\end{figure}

The density of NI vertices before background subtraction, reconstructed in the BPIX detector (\AbsZbarrel)
in the region of the beam pipe ring, projected onto the $x$-$y$ plane,
is shown in Fig.~\ref{fig:BeamPipe_Draw}.
The section of the pipe observed in the figure is machined as a thin beryllium cylinder, 0.8\unit{mm} thick. The pipe is maintained by collars located at $z = \pm 1.5\unit{m}$, which are outside of the region that is investigated by this analysis. The structure is therefore modeled by a simple circle in the $x$-$y$ plane.  The combinatorial background appears in blue in the figure. Since the axis of the pipe is shifted by approximately 1\unit{\mm} with respect to the coordinate system, we consider the whole region between $\rhomin = 2.0$ and $\rhomax = 2.4\unit{cm}$ for the fit.

\begin{figure}
\centering
\includegraphics[width=0.6\textwidth]{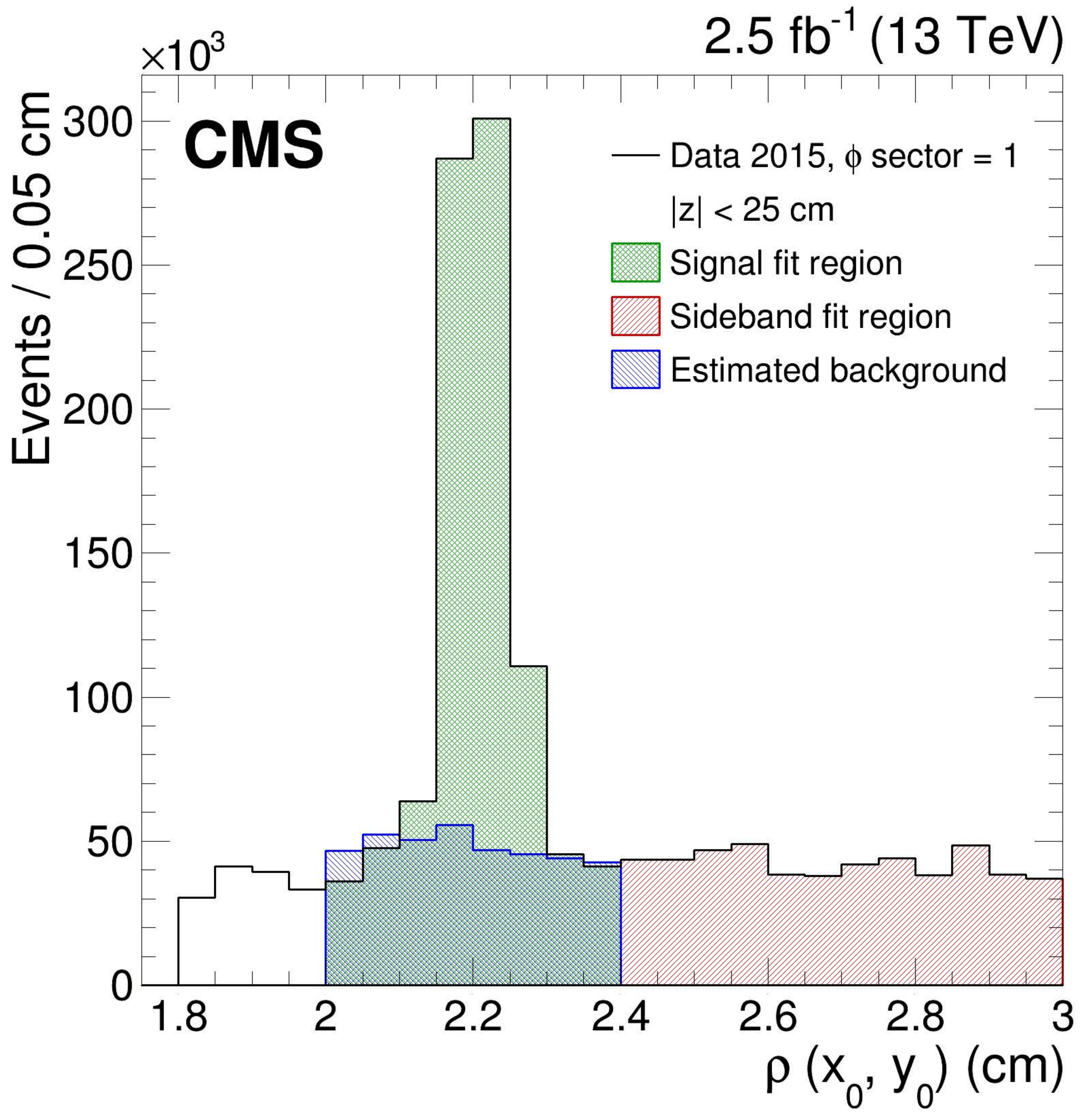}
\caption{The density of NI vertices versus \rhoZero for a $\phi$ slice of the beam pipe located near $\phi = 0$ (black line)
for \AbsZbarrel before background subtraction.
The green hatched area corresponds to the signal region,
the red hatched area corresponds to the sideband  region used to fit the background, and
the blue hatched area corresponds to the estimated background in the signal region.
}
\label{fig:BeamPipe_Slice}
\end{figure}

An example of a $\phi$ slice is shown in Fig.~\ref{fig:BeamPipe_Slice}. We clearly see a peak at around $\rhoZero = 2.25\unit{cm}$
that represents the location of the beam pipe.
The combinatorial background under the peak is estimated from the right sideband defined by $ 2.4 < \rhoZero < 3.0\unit{cm}$.  An exponential function is fitted to the sideband region and extrapolated into the signal region.

\begin{figure}
\centering
\includegraphics[width=0.6\textwidth]{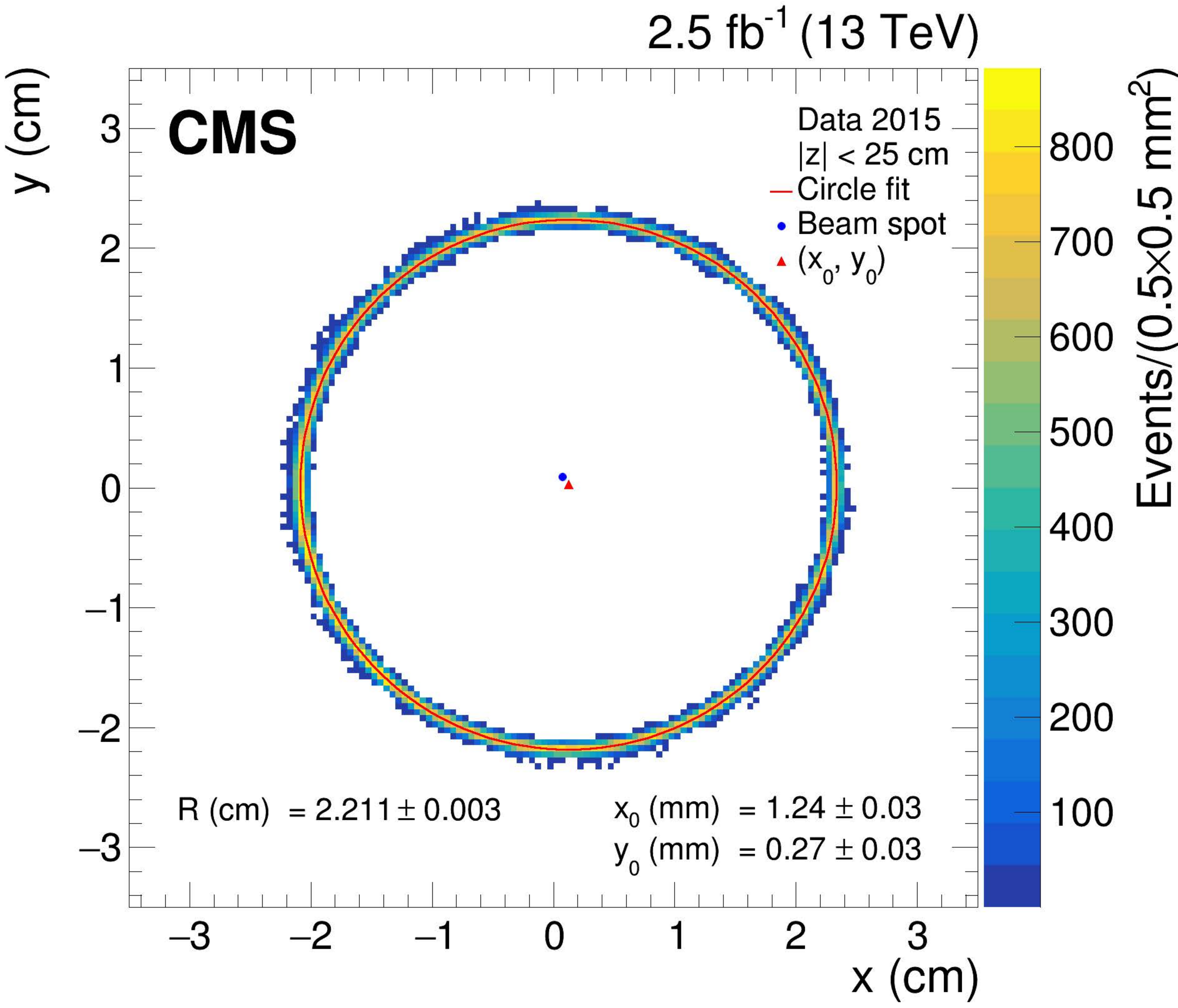}
\includegraphics[width=0.6\textwidth]{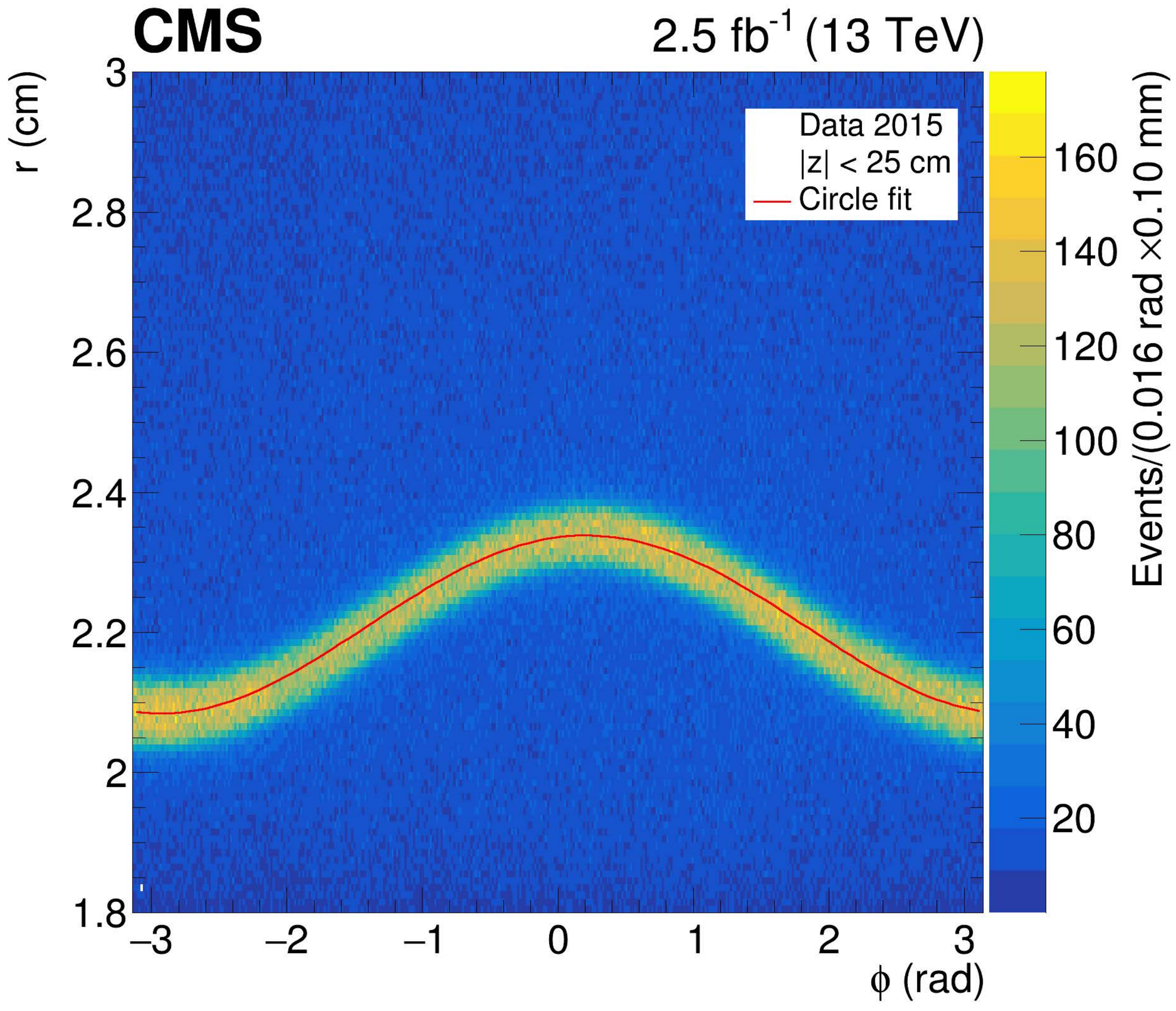}
\caption{The beam pipe region with the fitted values for a circle of radius \rc and center$(\xz, \yz)$ for \AbsZbarrel.
The $x$-$y$ plane after background subtraction (upper), and the $r$-$\phi$ coordinates before background subtraction (lower), are shown.
The density of NI vertices is indicated by the color scale.
The red line shows the fitted circle.
The blue point in the center of the $x$-$y$ plane corresponds to the average beam spot position
of $\xbs=0.8\unit{mm}$ and $\ybs=0.9\unit{mm}$ in 2015.}
\label{fig:BeamPipe_Fit}
\end{figure}

In Fig.~\ref{fig:BeamPipe_Fit}, the fit results for a circle of radius \rc and center $(\xz, \yz)$
are shown in the $x$-$y$ plane (upper), and $r$-$\phi$ coordinates (lower).
The radius is measured with a precision of 30\unit{\mum}, well below the thickness of the beam pipe. The radius measurement matches exactly the design value of the beam pipe, which is 22.1\unit{mm}~\cite{Dominguez:1481838}.
The center of the beam pipe is shifted by 1.24\unit{mm} in $x$ and 0.27\unit{mm} in $y$. The effect of this shift is visible in the sinusoidal modulation of the $r$-$\phi$ distribution, which is well modeled by the fit.

\subsection{Measurement of the BPIX detector inner shield position}\label{sec:PixelShield}

\begin{figure}
\centering
\includegraphics[width=0.6\textwidth]{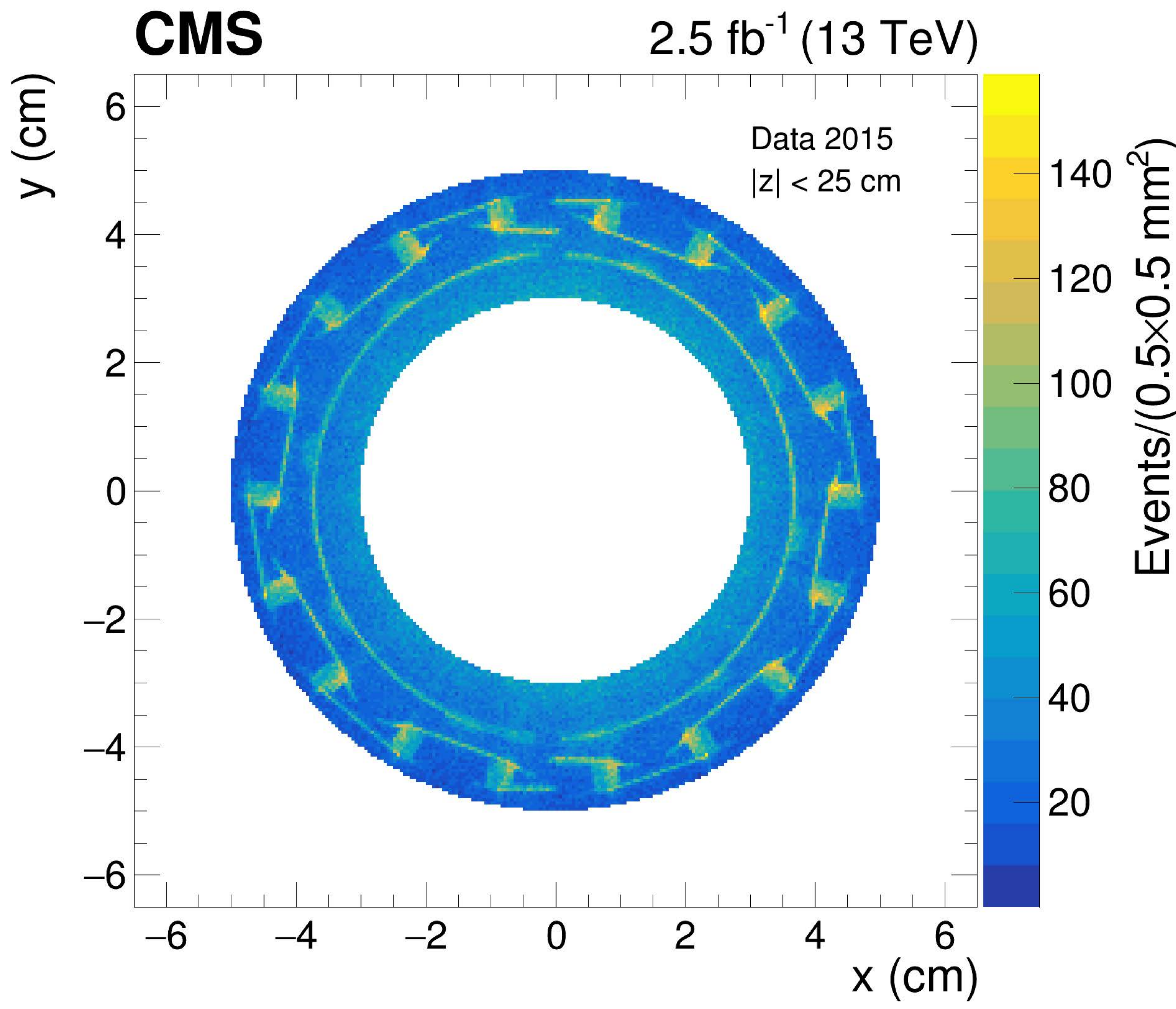}
\caption{The BPIX detector inner shield region viewed in the $x$-$y$ plane for \AbsZbarrel
before background subtraction and removal of the $\phi$ regions with additional structures.
The density of NI vertices is indicated by the color scale. The inner shield itself is
the visible circle of radius
$r = 3.8\unit{cm}$.
Modules in the first BPIX detector layer are visible at larger radius.
The small bumps that can be seen around the shield correspond to cables connected to the first BPIX detector layer.}
\label{fig:PixelShield_Draw}
\end{figure}

Figure~\ref{fig:PixelShield_Draw} shows the density of reconstructed NI vertices 
in the region of the BPIX detector inner shield for \AbsZbarrel as projected onto the $x$-$y$ plane.
The inner shield can be identified at around $r = 3.8\unit{cm}$ and protects the sensitive modules
of the first BPIX detector layer visible in the region around $r = 4\unit{cm}$.
The small bumps around the shield that are visible in Fig.~\ref{fig:PixelShield_Draw} correspond to the cables
connected to the first BPIX detector layer. The twelve $\phi$ sectors
that contain the cables are removed from the fit described below.

The background for the remaining $\phi$ sectors is estimated from the left sideband defined by $ 3.00 < \rhoZero < 3.55\unit{cm}$.
This region between the beam pipe and the BPIX detector inner shield is empty of any structure,
while the region between the inner shield and first BPIX detector layer is very small and occupied by cables.
The signal-over-background ratio for the BPIX detector inner shield is less favorable than for the beam pipe because the shield has a smaller amount of material.

\begin{figure}
\centering
\includegraphics[width=0.6\textwidth]{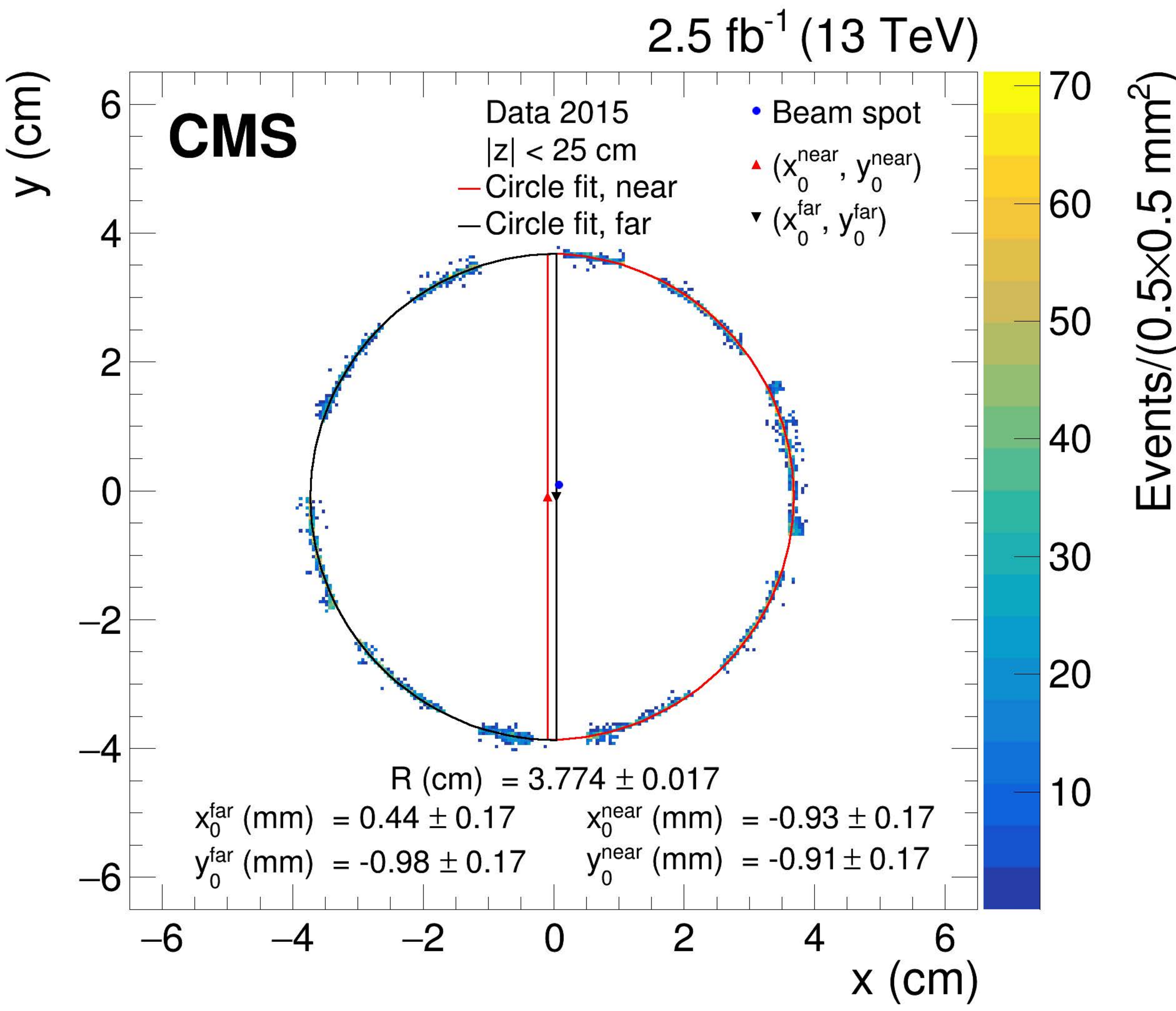}
\includegraphics[width=0.6\textwidth]{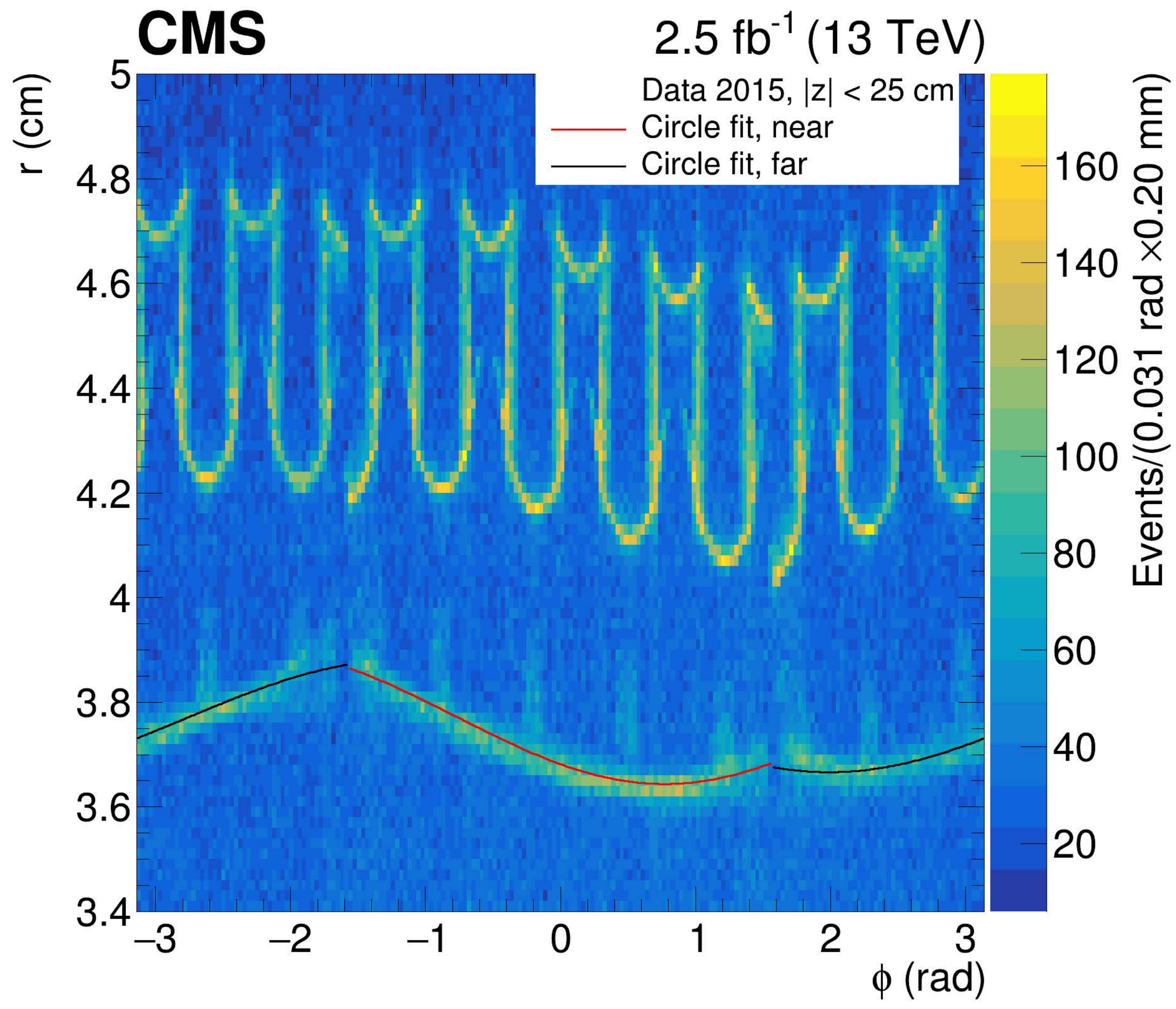}
\caption{The BPIX detector inner shield  with the fitted values for two half-circles of common radius \rc
and centers (\xZfar, \yZfar) and (\xZnear, \yZnear) for \AbsZbarrel.
The $x$-$y$ plane after background subtraction (upper), and the $r$-$\phi$ coordinates before background subtraction (lower), are shown.
The density of NI vertices is indicated by the color scale.
The red and black lines at around $r = 3.8\unit{cm}$ show the fitted half-circles on the far and near sides, respectively.
The blue point at the center of the $x$-$y$ plane corresponds to the average beam spot position
of $\xbs=0.8\unit{mm}$ and $\ybs=0.9\unit{mm}$ in 2015.
Modules in the first BPIX detector layer are visible in~(lower)  at larger radius.
}
\label{fig:PixelShield_Fit}
\end{figure}

In Fig.~\ref{fig:PixelShield_Fit}, the result of the fit to the BPIX detector inner shield with two half-circles
is shown in the $x$-$y$ plane (upper), and $r$-$\phi$ coordinates (lower).
The radii of the halves are assumed to be equal and represented by the parameter \rc. The possibilities that the radii could be different and that we have two half-ellipses instead of circles were tested and represent the dominant systematic uncertainty, which amounts to 170\unit{\mum}. The centers of each half-circle,
(\xZfar, \yZfar) and (\xZnear, \yZnear), are determined from the fit. The fit values for \yZfar and \yZnear show that the halves are vertically aligned to within 100\unit{\mum}.
The geometric shapes of the two half-circles used in the fit overlap, as seen in Fig.~\ref{fig:PixelShield_Fit}~(upper).
However, each half of the actual BPIX detector inner shield spans less than a half-circle in arc length,
resulting in a small gap between the halves that can be seen in Figs.~\ref{fig:PixelShield_Draw} and~\ref{fig:PixelShield_Fit}~(lower).

\subsection{Measurements of the positions of the pixel detector support tube and the BPIX detector support rails}\label{sec:PixelSupport}

\begin{figure}
\centering
\includegraphics[width=0.6\textwidth]{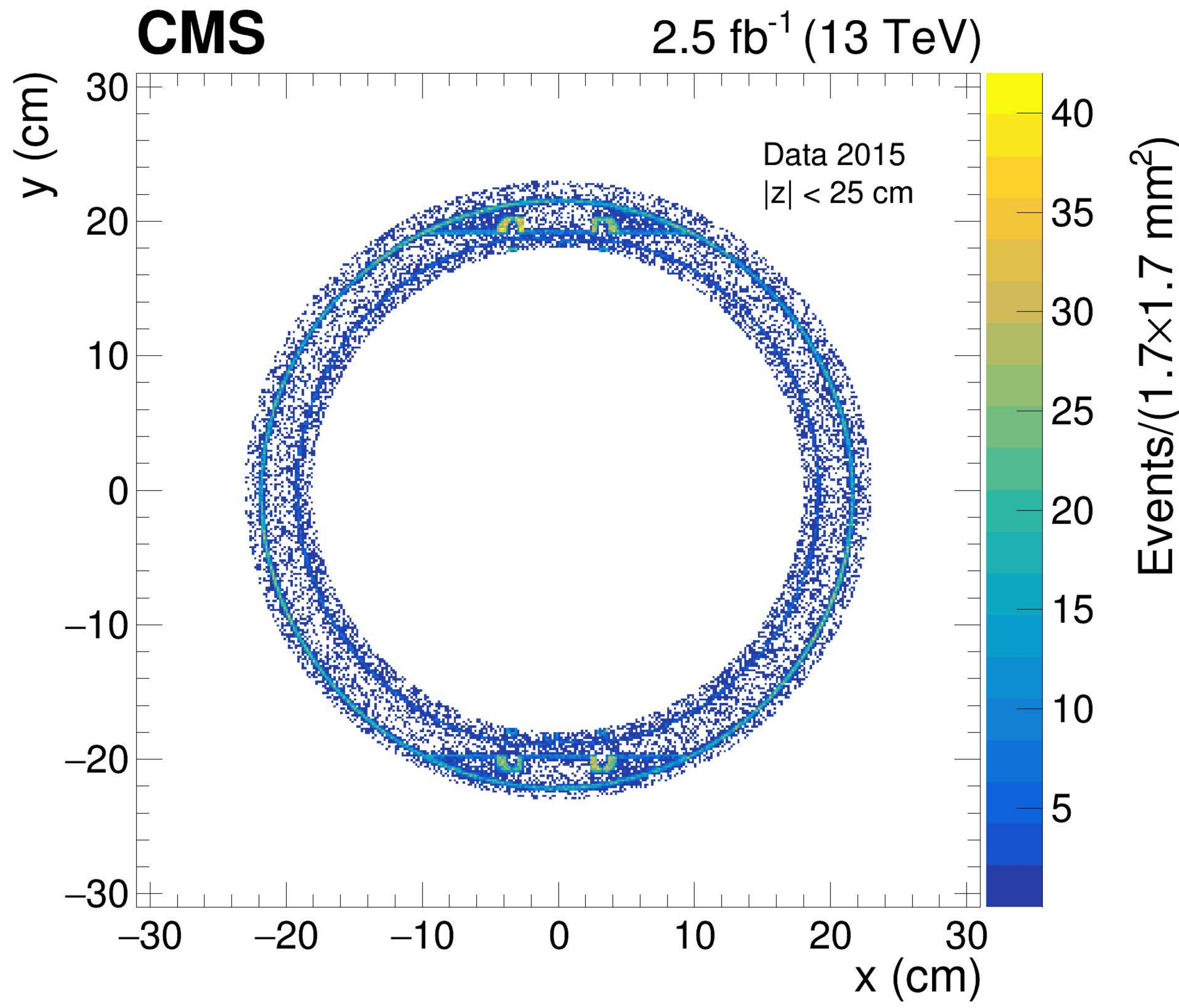}
\caption{The region of the pixel detector support tube viewed in the $x$-$y$ plane for \AbsZbarrel
before background subtraction and removal of the $\phi$ regions with additional structures.
The density of NI vertices is indicated by the color scale.
Two circular structures are visible.
The circle with the smaller radius corresponds to the BPIX detector outer shield,
while the one with the larger radius is the pixel detector support tube (also visible in Fig.~\ref{fig:TrackerHadrography}).}
\label{fig:PixelSupport_Draw}
\end{figure}

The density of NI vertices, reconstructed in the barrel tracking detector (\AbsZbarrel),
in the region of the pixel detector support tube, projected onto the $x$-$y$ plane,
is shown in Fig.~\ref{fig:PixelSupport_Draw}. The BPIX detector is placed within the cylinder of the pixel detector support tube using the top and bottom rails visible at $y \approx \pm 19\unit{cm}$.
The region around the rails (twelve $\phi$ sectors) is removed from the fitting procedure for the pixel detector support tube.

\begin{figure}
\centering
\includegraphics[width=0.6\textwidth]{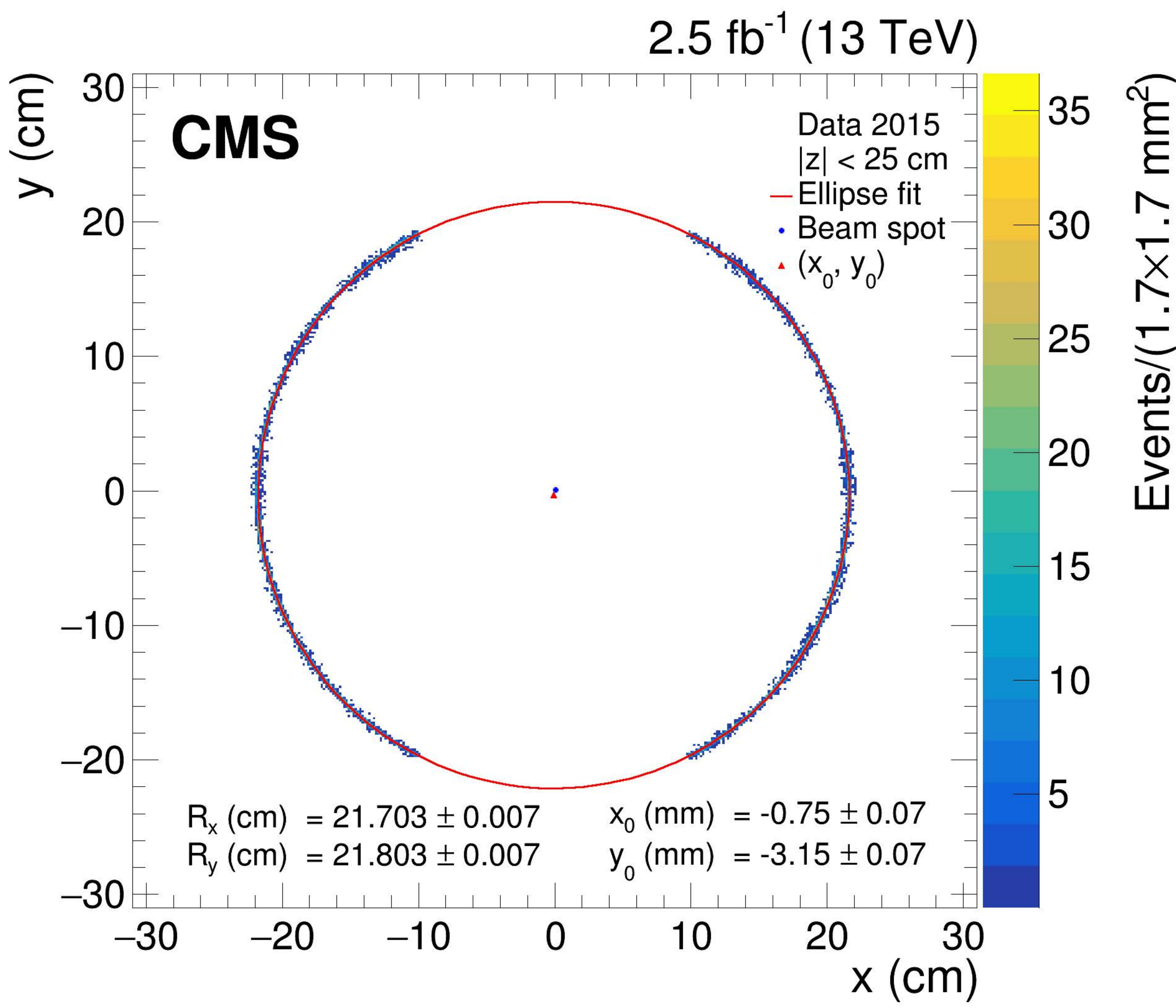}
\includegraphics[width=0.6\textwidth]{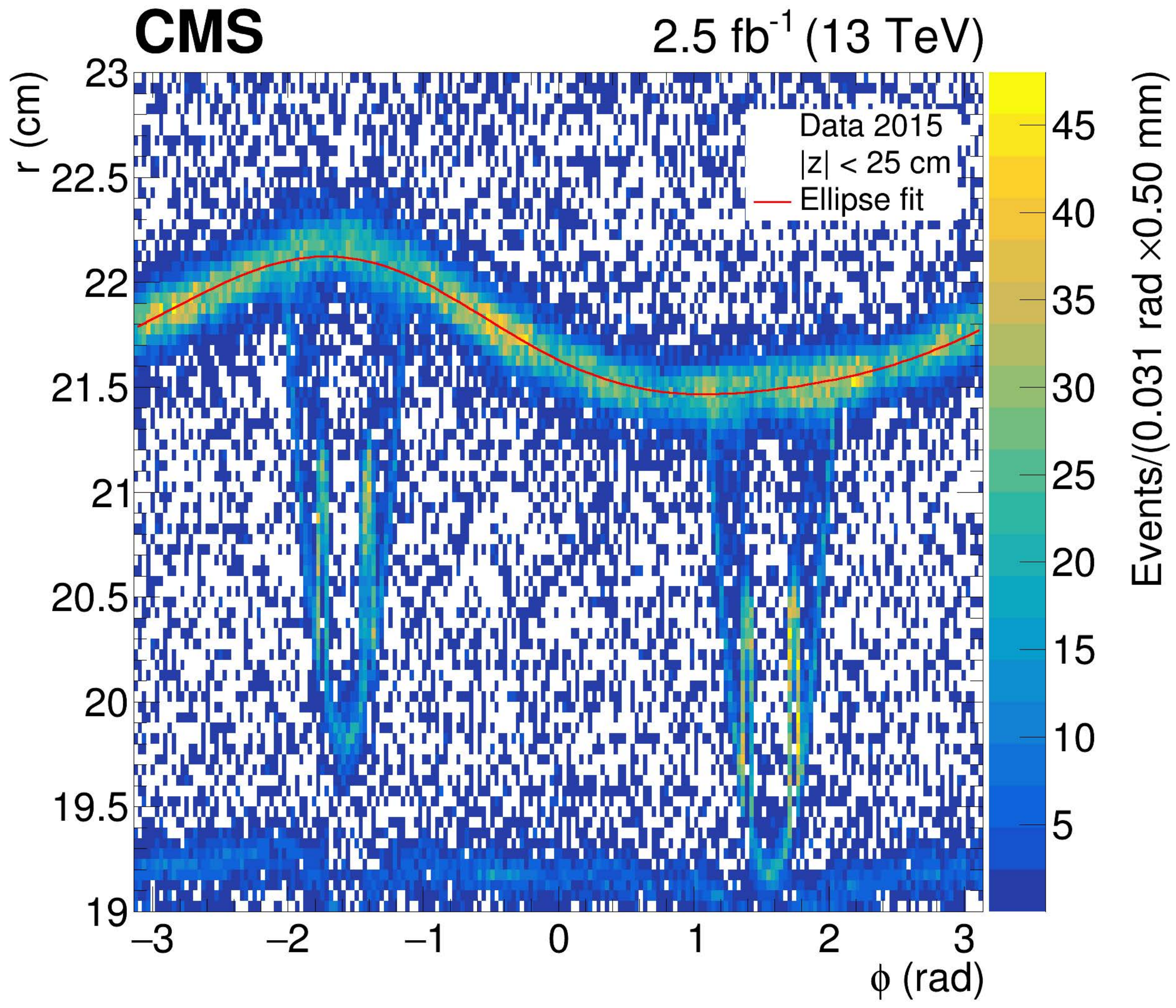}
\caption{The pixel detector support tube with the fitted values for an ellipse with semi-minor axis \Rx, semi-major axis \Ry,
and center $(\xz,\yz)$ for \AbsZbarrel.
The $x$-$y$ plane after background subtraction (upper), and the $r$-$\phi$ coordinates before background subtraction (lower), are shown.
The density of NI vertices is indicated by the color scale.
The red line shows the fitted ellipse.
The blue point in the center of the $x$-$y$ plane corresponds to the average beam spot position
of $\xbs=0.8\unit{mm}$ and $\ybs=0.9\unit{mm}$ in 2015.}
\label{fig:PixelSupport_Fit}
\end{figure}

When the pixel detector support tube was manufactured, it was circular, but it was deformed into an ellipse under its own weight after installation.
The pixel detector support tube was fitted with an ellipse because of this deformation, and a difference of 1.0\unit{mm} is seen between the two semi-axes.
In Fig.~\ref{fig:PixelSupport_Fit}, the result of the fit to the pixel detector support tube is shown
in the $x$-$y$ plane (upper), and $r$-$\phi$ coordinates (lower).
The semi-minor axis appears to be aligned with the $x$ axis and the semi-major axis with the $y$ axis.
The position of the center of the pixel detector support tube is shifted by a few millimeters with respect to the coordinate system.

\begin{figure}
\centering
\includegraphics[width=0.6\textwidth]{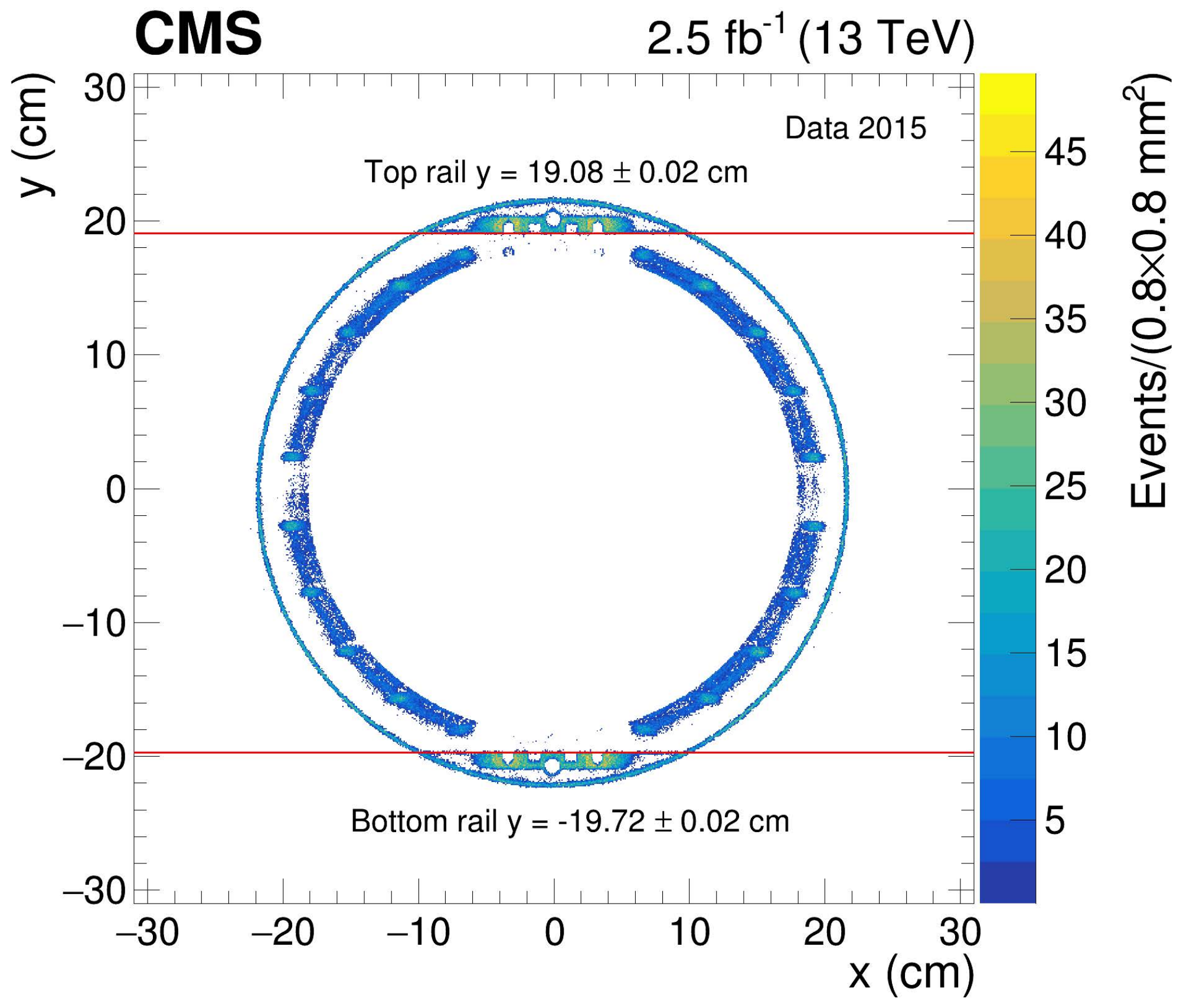}
\caption{
The BPIX detector support rails after background subtraction in the $x$-$y$ plane for the combined tracking detector barrel and endcap regions.
Horizontal red lines correspond to the fit of the BPIX detector support rails.
The density of NI vertices is indicated by the color scale.}
\label{fig:PixelSupportRails_Fit}
\end{figure}

The method used to measure the positions of the BPIX detector support rails is different than for the other inactive elements
since the rails are more complex structures to model.
The rails are mounted on support structures that are aligned with the $x$ axis, therefore we can identify the $y$ coordinate
(top rail \yr and bottom rail \yr) of this support structure.
In practice we define top and bottom rail \yr as the inner coordinate of the support structure,
estimated by finding the position with the maximum $y$ derivative in Fig.~\ref{fig:PixelSupportRails_Fit}.
The measurement is performed separately for the $y < 0$ and $y>0$ sides. Since the support structure is very thin, it is included in a single
bin of width 800\unit{\mum}. 
The uncertainty for a uniform/flat distribution is 
$1/\sqrt{12}$ of the bin size, \ie, 0.02\unit{\cm}.
This uncertainty includes effects from the fitting procedure and from small structures within the rails.
The results are shown in Fig.~\ref{fig:PixelSupportRails_Fit}
for the combined tracking detector barrel and endcap regions.
We also performed separate measurements for the barrel and endcap regions,
and the results were consistent with those obtained from the combined regions.

\subsection{Results}\label{sec:Results}

Table~\ref{table:Results} summarizes the results of the fits. The values of the parameters are tabulated for the fits to the beam pipe with a circle,
the BPIX detector inner shield with two half-circles, and the pixel detector support tube with an ellipse.
Only systematic uncertainties are provided since the statistical uncertainties are negligible (below 10\unit{\mum}).
Table~\ref{table:ResultsRails} summarizes the final results where the BPIX detector support rails were fitted with a horizontal line.
As an estimate of the systematic uncertainty we take half the bin width in $y$; the statistical uncertainties are once again negligible.

\begin{table}[h]
\centering
\topcaption {Results of the fit to the beam pipe with a circle, the BPIX detector inner shield with two half-circles,
and the pixel detector support tube with an ellipse.
Only systematic uncertainties are provided since the statistical uncertainties are negligible. }
\begin{tabular}{llll}
\hline
Object            & R (cm)             & \xz (mm)         & \yz (mm) \\
\hline
Beam pipe         & $2.211\pm0.003$    & $1.24\pm0.03$    & $0.27\pm0.03$ \\
BPIX detector inner shield, far  & $3.774\pm0.017$    & $0.44\pm0.17$    & $-0.98\pm0.17$ \\
BPIX detector inner shield, near & $3.774\pm0.017$    & $-0.93\pm0.17$   & $-0.91\pm0.17$ \\
Pixel detector support tube    & \Rx: $21.703\pm0.007$ & $-0.75\pm0.07$ & $-3.15\pm0.07$ \\
                  & \Ry: $21.803\pm0.007$ & & \\
\hline
\end{tabular}
\label{table:Results}
\end{table}

\begin{table}[h]
\centering
\topcaption {
Results of the fitted $y$ coordinate of the bottom and top BPIX detector support rails with a horizontal line.
Only systematic uncertainties are provided since the statistical uncertainties are negligible. }
\begin{tabular}{ll}
\hline
Bottom rail \yr (cm) & Top rail \yr (cm) \\
\hline
$-19.72\pm0.02$ & $19.08\pm0.02$ \\
\hline
\end{tabular}
\label{table:ResultsRails}
\end{table}

\section{Systematic uncertainties}\label{sec:Systematics}

The precision of the measurements presented in Table~\ref{table:Results} is determined by the systematic uncertainties.
These uncertainties are associated with the procedures for the background subtraction, the shape assumptions, parameter fitting, and NI vertex reconstruction.

\begin{table}[ht]
\centering
\topcaption {Systematic uncertainties in the position and radius measurements of three inactive detector elements.}
\begin{tabular}{llll}
\hline
Source of systematic uncertainty &  Beam pipe      & BPIX detector       & Pixel detector\\
                                 & (mm)            & inner shield (mm)   & support tube (mm) \\
\hline
Background shape       & 0.02        & 0.02               & 0.04 \\
\qquad variation & & & \\
Background subtraction    & $<$0.01       & 0.06               & 0.01 \\
Average background        & $<$0.01      & 0.02           & $<$0.01 \\
\qquad in three neighbor $\phi$ bins & & & \\
Structure shape             & $<$0.01        & 0.07 (position)     & \NA  \\
\qquad variation                &              & 0.14 (\Rnear,\Rfar) &  \\
Fit procedure       & 0.01            & 0.01               & 0.04 \\
Vertex reconstruction resolution         & 0.01         & 0.01               & 0.04 \\
\hline
Total                    & 0.03         & 0.17               & 0.07 \\
\hline
\end{tabular}
\label{table:Systematics}
\end{table}

Uncertainties arising from the subtraction and estimation of the background are determined by varying the shape and normalization of the background, and refitting the resulting signal. 
As a cross-check, instead of using an exponential fit, a simple horizontal line is fit in the sideband region and extrapolated into the signal region to determine the combinatorial background under the signal peak.
The largest difference seen in the fitted values for each structure is taken as the systematic uncertainty due to the background shape.  The normalization of the background was also varied.
The number of background events in each bin was
varied
by two standard deviations in the statistical uncertainty
and the resulting signal was refitted. The maximum difference in the fitted values is taken as the systematic uncertainty associated with the estimated size of the background subtraction.
A third background variation was performed by using the background estimated from neighboring $\phi$ bins.
Again, the difference in fitted values is taken as a systematic uncertainty. The uncertainties from these variations on the background are listed in Table~\ref{table:Systematics}. The combined effect of these three sources does not exceed 60\unit{\mum}.

The uncertainties in the measurements that were introduced by the assumptions made about the shapes of the beam pipe and BPIX detector inner shield
are also estimated
by refitting the data using different shapes.
The uncertainties in the shapes of the beam pipe and BPIX detector inner shield are estimated by fitting the beam pipe with an ellipse, instead of a circle,
and the BPIX detector inner shield with two half-ellipses, instead of two half-circles.
In the case of the beam pipe, it is observed that the two semi-axes of the ellipse are equal to within 10\unit{\mum}, which supports the use of a circular shape, and this difference is added to the systematic uncertainty of the radius measurement. The value of \Rx was fixed for the BPIX detector inner shield fits to the half-ellipse, because otherwise the fit was not stable. The results from the ellipse and half-ellipse fits were compared with the results from the circle and half-circle fits. The largest difference was taken as the systematic uncertainty from the shape of the structure.
The pixel detector support tube is already modeled with an ellipse,
and therefore no extra systematic uncertainty is assigned to the modeling of its shape.
The systematic uncertainties determined by varying the assumed shapes of the structures are listed in Table~\ref{table:Systematics}.

The systematic uncertainties associated with the fit procedure are determined using
pseudo-data
where the shapes and positions of the objects are fixed from the experimentally measured values.
Uncertainties for the fits to a circular shape are measured by generating a beam pipe centered at $(0,0)$, a beam pipe centered at the measured $(x,y)$, a BPIX detector inner shield generated with the values measured from the near side, and a BPIX detector inner shield generated with the values measured from the far side. The largest difference between the input parameters and the fit results was taken as the systematic uncertainty for the circle fit. The systematic uncertainty from the fit to an elliptical shape was determined similarly. Two pixel detector support tubes were generated. One was centered at $(0,0)$ with the semi-axes similar to the measured values, and the other was centered at the measured center of the pixel detector support tube with semi-axes similar to the measured values. The uncertainties found from fitting the simulated data are listed in Table~\ref{table:Systematics} in the ``Fit procedure" row.

Another source of systematic uncertainty comes from the reconstruction of the secondary vertex position for the NI.
The finite position resolution of the reconstructed vertices and the fitting procedure itself may introduce biases
in the position measurements. The effects of these potential biases are estimated by
measuring the structure properties in MC simulations based on single pions generated
where a cylindrical model is assumed for the beam pipe, BPIX detector inner shield and pixel detector support tubes centered on the beam axis. Pions with momenta
of 10 and 100\GeV are simulated. The simulated beam pipe, BPIX detector inner shield, and pixel detector support tube are centered at $(0,0)$.
The largest deviations from $(0,0)$ in the fits are taken as the systematic uncertainties, and are presented in Table~\ref{table:Systematics}
in the ``Vertex reconstruction resolution" row.

The systematic uncertainty in the position measurements for the BPIX detector support rails is conservatively estimated to be
$1/\sqrt{12}$ of the bin size.

During the 2015 data taking, CMS had cooling problems with its magnet, resulting in the magnet being cycled on and off several times. Since the changes in magnetic field could potentially affect the position of the beam pipe, the data were split into two halves chronologically to see if the position in the later data differed from that in the earlier data.
Within the measurement uncertainties, no change in position was observed.

\section{Comparison with technical surveys}\label{sec:Survey}

After the installation of the new CMS central beam pipe~\cite{Gallilee:1470582} during the 2013--2015 LHC shutdown, a number of measurements were taken in order to better understand the position and stability under different supporting configurations of the central beam pipe itself and later also of the BPIX detector.
The pixel detector support tube was not surveyed at this time because NI measurements had shown no motion of it with respect to earlier surveys and it is not possible to adjust its position in any case.

\subsection{CMS survey coordinate system}
\label{CMScoordinate}
The CMS coordinate system used by the surveyors is the same as that described in Section~2.
The local geometry of the CMS cavern is a local transformation from the global geometry of the LHC.
The coordinate system used for offline analysis in CMS is based on a 3D best fit reconstruction from high \pt tracks coming from the interaction region of the TOB centroid. The reconstructed TOB centroid defines the CMS detector central axis and, based on measurements taken by the surveyors at CMS after the installation of the tracking system in 2009 in the CMS coordinate system, is made to coincide with the latter in the offline code via simple rigid translations and rotations.
The two coordinate systems (CMS offline system and CMS coordinate system) should then coincide within the uncertainties that are dominated by the surveyors measurements. The systematic uncertainty is estimated
to be $\pm$0.75\unit{mm} and it should be added to any other quoted uncertainties when comparing results obtained within the two systems.

\subsection{Central beam pipe}

The CMS central beam pipe spans the central detector region over 6.2\unit{m} and is held vertically and horizontally by metal wires attached to collars positioned at $\pm$1.6\unit{m} from the center and at the two extremities ($\pm$3.1\unit{m}) by flanges connected to the endcap sections of the beam pipe.

Measurements of the position of the beam pipe were taken using a theodolite and measuring four points for each collar and flange. A best fit to a circle gives the position of the center of the beam pipe in the four positions along the $z$ axis. The accuracy of this measurement is estimated to be $\pm$0.5\unit{mm}.
Table~\ref{table:survey_pipe} summarizes the results of the
measurements of the CMS central beam pipe on January 12, 2015 after the re-installation of
the original pixel detector (both barrel and endcaps parts) and with all supports in their final configuration.

\begin{table}[h]
\begin{center}
\topcaption {Results from the survey of the CMS central beam pipe positions on January 12, 2015.}
\begin{tabular}{llll}
\hline
Support           & $x$ (\unit{mm})            & $y$ (\unit{mm})         & $z$ (\unit{mm}) \\
\hline
Flange $+z$ ($+3.1\unit{m}$)  & 0.2    & 0.2   & 3131.7 \\
Collar $+z$ ($+1.6\unit{m}$)  & 0.8    & 0.2   & \NA \\
Collar $-z$ ($-1.6\unit{m}$)  & 0.7    & 0.1   & \NA \\
Flange $-z$ ($-3.1\unit{m}$)  & $-$0.8    & 0.2  & $-$3136.9 \\
\hline
\end{tabular}
\label{table:survey_pipe}
\end{center}
\end{table}

Although these results should be directly compared with the NI measurements taken shortly thereafter during 2015 data taking, there are several considerations to be made, potentially leading to a somewhat different position of the beam pipe during data taking:

\begin{itemize}
\item The CMS endcaps needed to be closed, and in the process, various beam pipe supports are temporarily removed and exchanged.
\item The magnetic field was turned on, compressing the endcaps inside the solenoid toward the interaction point (hence the need for various supports along the beam pipe).
\item Vacuum was created inside the beam pipe before the beam can circulate.
\item The ambient temperature in the tracking system and central beam pipe volume went down to around 0\de\unit{C} during data taking.
\end{itemize}

Notwithstanding these differences, the survey coordinates of the center of the beam pipe are compatible with the NI measurements. 
The $x$ and $y$ coordinates in Table~\ref{table:Results} for \AbsZbarrel agree within uncertainties 
with the inner beam pipe (collar) positions given in Table~\ref{table:survey_pipe}.

\subsection{BPIX detector}

In order to be able to measure the position of the BPIX detector right after installation, optical targets were glued onto the end flanges of the detector,
which are visible from outside the pixel detector support tube using a theodolite positioned on the pixel detector installation platform at each end. The BPIX detector is divided into
two separate parts
labeled far and near
as discussed in Section~\ref{sec:CMSdetector}.
The detector itself spans the interaction region over about 50\unit{cm} in the $z$ direction,
but the services, running along the outside surface of the pixel detector supply tube, extend to the end of the pixel detector volume at $z = \pm 3.0\unit{m}$.

Three survey target points, indicated by the numbers 2001, 2002, and 2003, are visible in Fig.~\ref{fig:PhotoHalve}~(right): one survey mark and two mechanical flat screws were used to
define the plane of the end flange of the detector,
and the positions are measured
using photogrammetry techniques in the laboratory before installation. The three points on each of the four end flanges of the detector
(near $+z$, near $-z$, far $+z$, and far $-z$)
were then referenced to the center axis of the BPIX detector with an estimated accuracy of $\pm$0.2\unit{mm}.
The detector was installed on December 11, 2014 and its position measured with the theodolite. The three points at each end define a plane and a center, and the two planes combined define a center line in 3D space. Each side (far and near) is treated independently yielding two center lines in 3D space,
one extrapolated from measurements of the far side and one from measurements of the near side. The accuracy of the two extrapolated center lines is estimated to be $\pm$1.0\unit{mm}.
From these measurements the survey determined that the overall detector center is low by 1.1\unit{mm} ($y=-1.1\unit{mm}$
for both the far and near halves), in good agreement with the \yz results shown in Table~\ref{table:Results}.
In the $x$ direction, the average of the far and near center positions is $-$0.7\unit{mm} in the survey, which agrees within the uncertainties with the
average value of $-$0.2\unit{mm} from the NI measurements.

\section{Summary}\label{sec:Conclusions}

Nuclear interactions have a reputation of being undesirable events that degrade the quality of the reconstruction of charged and neutral hadrons.
In this analysis, it has been demonstrated that they can be used to produce a high-precision map of the material inside the tracking system. 
Such maps can be useful for validating detector simulations and identifying any shifts in detector elements during operation.

Using a data set that corresponds to an integrated luminosity of 2.5\fbinv of proton-proton collisions at a center-of-mass energy of 13\TeV, a large sample of secondary hadronic interactions was collected. After background subtraction, the positions of the secondary vertices were used to determine the locations of inactive elements with a precision of the order of 100\unit{\mum}.
The positions of the beam pipe and the inner tracking system structures
(pixel detector support tube, and BPIX inner shield and support rails)
were determined with a precision that depends on the structure under study.
No significant position bias was identified through the technique, and statistical uncertainties were negligible.
The positions of the structures under consideration were probed with a precision better than the typical installation tolerances and are found to be compatible with previous survey measurements.

\begin{acknowledgments}
We congratulate our colleagues in the CERN accelerator departments for the excellent performance of the LHC and thank the technical and administrative staffs at CERN and at other CMS institutes for their contributions to the success of the CMS effort. In addition, we gratefully acknowledge the computing centers and personnel of the Worldwide LHC Computing Grid for delivering so effectively the computing infrastructure essential to our analyses. Finally, we acknowledge the enduring support for the construction and operation of the LHC and the CMS detector provided by the following funding agencies: BMWFW and FWF (Austria); FNRS and FWO (Belgium); CNPq, CAPES, FAPERJ, FAPERGS, and FAPESP (Brazil); MES (Bulgaria); CERN; CAS, MoST, and NSFC (China); COLCIENCIAS (Colombia); MSES and CSF (Croatia); RPF (Cyprus); SENESCYT (Ecuador); MoER, ERC IUT, and ERDF (Estonia); Academy of Finland, MEC, and HIP (Finland); CEA and CNRS/IN2P3 (France); BMBF, DFG, and HGF (Germany); GSRT (Greece); NKFIA (Hungary); DAE and DST (India); IPM (Iran); SFI (Ireland); INFN (Italy); MSIP and NRF (Republic of Korea); LAS (Lithuania); MOE and UM (Malaysia); BUAP, CINVESTAV, CONACYT, LNS, SEP, and UASLP-FAI (Mexico); MOS (Montenegro); MBIE (New Zealand); PAEC (Pakistan); MSHE and NSC (Poland); FCT (Portugal); JINR (Dubna); MON, RosAtom, RAS, RFBR, and NRC KI (Russia); MESTD (Serbia); SEIDI, CPAN, PCTI, and FEDER (Spain); MOSTR (Sri Lanka); Swiss Funding Agencies (Switzerland); MST (Taipei); ThEPCenter, IPST, STAR, and NSTDA (Thailand); TUBITAK and TAEK (Turkey); NASU and SFFR (Ukraine); STFC (United Kingdom); DOE and NSF (USA).
\hyphenation{Rachada-pisek} Individuals have received support from the Marie-Curie program and the European Research Council and Horizon 2020 Grant, contract No. 675440 (European Union); the Leventis Foundation; the A. P. Sloan Foundation; the Alexander von Humboldt Foundation; the Belgian Federal Science Policy Office; the Fonds pour la Formation \`a la Recherche dans l'Industrie et dans l'Agriculture (FRIA-Belgium); the Agentschap voor Innovatie door Wetenschap en Technologie (IWT-Belgium); the F.R.S.-FNRS and FWO (Belgium) under the ``Excellence of Science - EOS" - be.h project n. 30820817; the Ministry of Education, Youth and Sports (MEYS) of the Czech Republic; the Lend\"ulet (``Momentum") Programme and the J\'anos Bolyai Research Scholarship of the Hungarian Academy of Sciences, the New National Excellence Program \'UNKP, the NKFIA research grants 123842, 123959, 124845, 124850 and 125105 (Hungary); the Council of Science and Industrial Research, India; the HOMING PLUS program of the Foundation for Polish Science, cofinanced from European Union, Regional Development Fund, the Mobility Plus program of the Ministry of Science and Higher Education, the National Science Center (Poland), contracts Harmonia 2014/14/M/ST2/00428, Opus 2014/13/B/ST2/02543, 2014/15/B/ST2/03998, and 2015/19/B/ST2/02861, Sonata-bis 2012/07/E/ST2/01406; the National Priorities Research Program by Qatar National Research Fund; the Programa Estatal de Fomento de la Investigaci{\'o}n Cient{\'i}fica y T{\'e}cnica de Excelencia Mar\'{\i}a de Maeztu, grant MDM-2015-0509 and the Programa Severo Ochoa del Principado de Asturias; the Thalis and Aristeia programs cofinanced by EU-ESF and the Greek NSRF; the Rachadapisek Sompot Fund for Postdoctoral Fellowship, Chulalongkorn University and the Chulalongkorn Academic into Its 2nd Century Project Advancement Project (Thailand); the Welch Foundation, contract C-1845; and the Weston Havens Foundation (USA);
the Hellenic Foundation for Research and Innovation, HFRI; the Fondazione Ing. Aldo Gini.
\end{acknowledgments}

\bibliography{auto_generated}

\cleardoublepage \appendix\section{The CMS Collaboration \label{app:collab}}\begin{sloppypar}\hyphenpenalty=5000\widowpenalty=500\clubpenalty=5000\input{TRK-17-001-authorlist.tex}\end{sloppypar}
\end{document}

%% file: TRK-17-001-authorlist.tex
\vskip\cmsinstskip
\textbf{Yerevan Physics Institute, Yerevan, Armenia}\\*[0pt]
A.M.~Sirunyan, A.~Tumasyan
\vskip\cmsinstskip
\textbf{Institut f\"{u}r Hochenergiephysik, Wien, Austria}\\*[0pt]
W.~Adam, F.~Ambrogi, E.~Asilar, T.~Bergauer, J.~Brandstetter, E.~Brondolin, M.~Dragicevic, J.~Er\"{o}, A.~Escalante~Del~Valle, M.~Flechl, R.~Fr\"{u}hwirth\cmsAuthorMark{1}, V.M.~Ghete, J.~Grossmann, J.~Hrubec, M.~Jeitler\cmsAuthorMark{1}, A.~K\"{o}nig, N.~Krammer, I.~Kr\"{a}tschmer, D.~Liko, T.~Madlener, I.~Mikulec, N.~Rad, H.~Rohringer, J.~Schieck\cmsAuthorMark{1}, R.~Sch\"{o}fbeck, M.~Spanring, D.~Spitzbart, H.~Steininger, A.~Taurok, W.~Waltenberger, J.~Wittmann, C.-E.~Wulz\cmsAuthorMark{1}, M.~Zarucki
% ...
\vskip\cmsinstskip
% ...
1:  Also at Vienna University of Technology, Vienna, Austria\\
% ...


%% file: TRK-17-001_temp.bbl
\providecommand{\href}[2]{#2}\begingroup\raggedright\begin{thebibliography}{10}%
\makeatletter
\providecommand{\hrefCMSnoop }[0]{\@secondoftwo}%
\makeatother
\providecommand{\doi}{\texttt{doi:}\begingroup \urlstyle{tt}\Url}

\bibitem{Chatrchyan:2008zzk}
\hrefCMSnoop {}{{CMS Collaboration}, ``The {CMS} experiment at the {CERN}
  {LHC}'',} \textit{ JINST} \textbf{ 3} (2008) S08004,
\href{http://dx.doi.org/10.1088/1748-0221/3/08/S08004}{\doi{10.1088/1748-0221/3/08/S08004}}.

\bibitem{Khachatryan:2016sfv}
\hrefCMSnoop {}{{CMS Collaboration}, ``Search for long-lived charged particles
  in proton-proton collisions at $\sqrt{s} = 13$ {TeV}'',} \textit{ Phys. Rev.
  D} \textbf{ 94} (2016) 112004,
  \href{http://dx.doi.org/10.1103/PhysRevD.94.112004}{\doi{10.1103/PhysRevD.94.112004}},
\href{http://www.arXiv.org/abs/1609.08382}{\texttt{arXiv:1609.08382}}.

\bibitem{Sirunyan:2017jdo}
\hrefCMSnoop {}{{CMS Collaboration}, ``Search for new long-lived particles at
  $\sqrt{s} = 13$ {TeV}'',} \textit{ Phys. Lett. B} \textbf{ 780} (2018) 432,
  \href{http://dx.doi.org/10.1016/j.physletb.2018.03.019}{\doi{10.1016/j.physletb.2018.03.019}},
\href{http://www.arXiv.org/abs/1711.09120}{\texttt{arXiv:1711.09120}}.

\bibitem{Sirunyan:2017ezt}
\hrefCMSnoop {}{{CMS Collaboration}, ``Identification of heavy-flavour jets
  with the {CMS} detector in pp collisions at 13 {TeV}'',} \textit{ JINST}
  \textbf{ 13} (2018) P05011,
  \href{http://dx.doi.org/10.1088/1748-0221/13/05/P05011}{\doi{10.1088/1748-0221/13/05/P05011}},
\href{http://www.arXiv.org/abs/1712.07158}{\texttt{arXiv:1712.07158}}.

\bibitem{Agostinelli:2002hh}
\hrefCMSnoop {}{{GEANT4} Collaboration, ``{\GEANTfour}---a simulation
  toolkit'',} \textit{ Nucl. Instrum. Meth. A} \textbf{ 506} (2003) 250,
\href{http://dx.doi.org/10.1016/S0168-9002(03)01368-8}{\doi{10.1016/S0168-9002(03)01368-8}}.

\bibitem{GEANT}
\hrefCMSnoop {}{J.~Allison {et~al.}, ``{\GEANTfour developments and
  applications}'',} \textit{ IEEE Trans. Nucl. Sci.} \textbf{ 53} (2006) 270,
\href{http://dx.doi.org/10.1109/TNS.2006.869826}{\doi{10.1109/TNS.2006.869826}}.

\bibitem{Migliore:2010cva}
\href {https://cds.cern.ch/record/1278158}{{CMS Collaboration}, ``Altered
  scenarios of the {CMS} tracker material for systematic uncertainties
  studies'',} Technical Report CMS-NOTE-2010-010, CERN, 2010.

\bibitem{CMS-PAS-TRK-10-003}
\href {http://cds.cern.ch/record/1279138}{{{CMS}} Collaboration, ``Studies of
  tracker material'',} CMS Physics Analysis Summary CMS-PAS-TRK-10-003, CERN,
  2010.

\bibitem{Khachatryan:2015iwa}
\hrefCMSnoop {}{{CMS Collaboration}, ``Performance of photon reconstruction and
  identification with the {CMS} detector in proton-proton collisions at
  $\sqrt{s} = 8$ {TeV}'',} \textit{ JINST} \textbf{ 10} (2015) P08010,
  \href{http://dx.doi.org/10.1088/1748-0221/10/08/P08010}{\doi{10.1088/1748-0221/10/08/P08010}},
\href{http://www.arXiv.org/abs/1502.02702}{\texttt{arXiv:1502.02702}}.

\bibitem{Dominguez:1481838}
\hrefCMSnoop {}{{CMS Collaboration}, ``{CMS} technical design report for the
  pixel detector upgrade'',} technical report, CERN, 2012.
\newblock \href{http://dx.doi.org/10.2172/1151650}{\doi{10.2172/1151650}}.

\bibitem{Aad:2011cxa}
\hrefCMSnoop {}{{ATLAS Collaboration}, ``A study of the material in the {ATLAS}
  inner detector using secondary hadronic interactions'',} \textit{ JINST}
  \textbf{ 7} (2012) P01013,
  \href{http://dx.doi.org/10.1088/1748-0221/7/01/P01013}{\doi{10.1088/1748-0221/7/01/P01013}},
\href{http://www.arXiv.org/abs/1110.6191}{\texttt{arXiv:1110.6191}}.

\bibitem{Aaboud:2016poq}
\hrefCMSnoop {}{{ATLAS Collaboration}, ``{A measurement of material in the
  ATLAS tracker using secondary hadronic interactions in 7 TeV pp
  collisions}'',} \textit{ JINST} \textbf{ 11} (2016) P11020,
  \href{http://dx.doi.org/10.1088/1748-0221/11/11/P11020}{\doi{10.1088/1748-0221/11/11/P11020}},
\href{http://www.arXiv.org/abs/1609.04305}{\texttt{arXiv:1609.04305}}.

\bibitem{Aaboud:2017pjd}
\hrefCMSnoop {}{{ATLAS Collaboration}, ``Study of the material of the {ATLAS}
  inner detector for {Run} 2 of the {LHC}'',} \textit{ JINST} \textbf{ 12}
  (2017) P12009,
  \href{http://dx.doi.org/10.1088/1748-0221/12/12/P12009}{\doi{10.1088/1748-0221/12/12/P12009}},
\href{http://www.arXiv.org/abs/1707.02826}{\texttt{arXiv:1707.02826}}.

\bibitem{Chatrchyan:2014wfa}
\hrefCMSnoop {}{{CMS Collaboration}, ``Alignment of the {CMS} tracker with
  {LHC} and cosmic ray data'',} \textit{ JINST} \textbf{ 9} (2014) P06009,
  \href{http://dx.doi.org/10.1088/1748-0221/9/06/P06009}{\doi{10.1088/1748-0221/9/06/P06009}},
\href{http://www.arXiv.org/abs/1403.2286}{\texttt{arXiv:1403.2286}}.

\bibitem{Pythia8}
\hrefCMSnoop {}{T.~Sj{\"o}strand, S.~Mrenna, and P.~Z. Skands, ``A brief
  introduction to {\PYTHIA} 8.1'',} \textit{ Comput. Phys. Commun.} \textbf{
  178} (2008) 852,
  \href{http://dx.doi.org/10.1016/j.cpc.2008.01.036}{\doi{10.1016/j.cpc.2008.01.036}},
\href{http://www.arXiv.org/abs/0710.3820}{\texttt{arXiv:0710.3820}}.

\bibitem{Sjostrand:2014zea}
T.~Sj{\"o}strand\hrefCMSnoop {}{ {et~al.}, ``An introduction to {\PYTHIA}
  8.2'',} \textit{ Comput. Phys. Commun.} \textbf{ 191} (2015) 159,
  \href{http://dx.doi.org/10.1016/j.cpc.2015.01.024}{\doi{10.1016/j.cpc.2015.01.024}},
\href{http://www.arXiv.org/abs/1410.3012}{\texttt{arXiv:1410.3012}}.

\bibitem{Khachatryan:2015pea}
\hrefCMSnoop {}{{CMS Collaboration}, ``Event generator tunes obtained from
  underlying event and multiparton scattering measurements'',} \textit{ Eur.
  Phys. J. C} \textbf{ 76} (2016) 155,
  \href{http://dx.doi.org/10.1140/epjc/s10052-016-3988-x}{\doi{10.1140/epjc/s10052-016-3988-x}},
\href{http://www.arXiv.org/abs/1512.00815}{\texttt{arXiv:1512.00815}}.

\bibitem{Khachatryan:2016bia}
\hrefCMSnoop {}{{CMS Collaboration}, ``The {CMS} trigger system'',} \textit{
  JINST} \textbf{ 12} (2017) P01020,
  \href{http://dx.doi.org/10.1088/1748-0221/12/01/P01020}{\doi{10.1088/1748-0221/12/01/P01020}},
\href{http://www.arXiv.org/abs/1609.02366}{\texttt{arXiv:1609.02366}}.

\bibitem{CMS_CR_2008_007}
{CMS Tracker} Collaboration, \hrefCMSnoop {}{R.~Ranieri, ``The simulation of
  the {CMS} silicon tracker'',} in \textit{ {Proceedings, 2007 IEEE Nuclear
  Science Symposium and Medical Imaging Conference (NSS/MIC): Honolulu,
  Hawaii}}, p.~2434.
\newblock 2007.
\newblock
\href{http://dx.doi.org/10.1109/NSSMIC.2007.4436649}{\doi{10.1109/NSSMIC.2007.4436649}}.

\bibitem{Chatrchyan:2014fea}
\hrefCMSnoop {}{{CMS Collaboration}, ``Description and performance of track and
  primary-vertex reconstruction with the {CMS} tracker'',} \textit{ JINST}
  \textbf{ 9} (2014) P10009,
  \href{http://dx.doi.org/10.1088/1748-0221/9/10/P10009}{\doi{10.1088/1748-0221/9/10/P10009}},
\href{http://www.arXiv.org/abs/1405.6569}{\texttt{arXiv:1405.6569}}.

\bibitem{Sirunyan:2017ulk}
\hrefCMSnoop {}{{CMS Collaboration}, ``Particle-flow reconstruction and global
  event description with the {CMS} detector'',} \textit{ JINST} \textbf{ 12}
  (2017) P10003,
  \href{http://dx.doi.org/10.1088/1748-0221/12/10/P10003}{\doi{10.1088/1748-0221/12/10/P10003}},
\href{http://www.arXiv.org/abs/1706.04965}{\texttt{arXiv:1706.04965}}.

\bibitem{CMS-DP-2016-012}
\href {https://cds.cern.ch/record/2155558}{{CMS Collaboration}, ``Tracking
  {POG} plot results on 2015 data'',} Technical Report CMS-DP-2016-012, CERN,
  2016.

\bibitem{Gallilee:1470582}
M.~Gallilee\href {https://cds.cern.ch/record/1470582}{ {et~al.}, ``{LHC}
  detector vacuum system consolidation for long shutdown 1 {(LS1)} in
  2013-2014'',} in \textit{ {Proceedings, 3rd International Conference on
  Particle accelerator (IPAC 2012): New Orleans, USA}}, p.~2555.
\newblock
2012.
\newblock

\end{thebibliography}\endgroup
